\documentclass[%
superscriptaddress,
nofootinbib,
twocolumn,
amsmath,amssymb,
aps,
prd,
floatfix,
a4paper
]{revtex4-2}

\usepackage[colorlinks=true,linkcolor=blue,citecolor=blue,linktocpage=false]{hyperref}

\usepackage{newtxtext,newtxmath}
\usepackage{graphicx}% Include figure files
\usepackage{dcolumn}% Align table columns on decimal point
\usepackage{bm}% bold math
\usepackage{color}
\usepackage{enumitem}
\usepackage{placeins}

\begin{document}

\preprint{}

\title{One-Loop Nonlinear Matter Power Spectrum from Unified Lagrangian Perturbation Theory: Fast Computation and Comparison with Emulators}

\author{Naonori Sugiyama}
\email{nao.s.sugiyama@gmail.com}
\affiliation{Independent Researcher, Tokyo, Japan}
\affiliation{National Astronomical Observatory of Japan, Mitaka, Tokyo 181-8588, Japan}%Lines break automatically or can be forced with \\
\thanks{Special Visiting Researcher (non-salaried)}

\date{\today}% It is always \today, today,
             %  but any date may be explicitly specified

\begin{abstract}

    We present a fast and accurate formulation for computing the nonlinear matter power spectrum at one-loop order based on Unified Lagrangian Perturbation Theory (ULPT). ULPT decomposes the density field into the Jacobian deviation, capturing intrinsic nonlinear growth, and the displacement-mapping factor, accounting for large-scale distortions due to bulk flows. This structural separation leads to a natural division of the power spectrum into a source term and a displacement-mapping factor, ensuring infrared (IR) safety by construction. We implement an efficient numerical algorithm using FFTLog and FAST-PT, achieving approximately 2-second evaluations on a standard laptop. The results are validated against simulation-based emulators, including the Dark Emulator and Euclid Emulator 2. Across 100 sampled cosmologies, ULPT agrees with emulator predictions at the 2--3\% level up to \( k \simeq 0.4\,h\,\mathrm{Mpc}^{-1} \) for \( z \geq 0.5 \), without any nuisance parameters. Similar agreement is found in configuration space, where the two-point correlation function remains accurate down to \( r \simeq 10\,h^{-1}\mathrm{Mpc} \). Compared to standard perturbation theory, which fails at small scales due to series expansion of the displacement factor, ULPT maintains convergence by preserving its full exponential form. We also clarify the mechanism of BAO damping: exponential suppression by displacement and peak sharpening by nonlinear growth. The combination accurately reproduces BAO features seen in simulations. ULPT thus offers a robust, IR-safe, and computationally efficient framework for modeling large-scale structure in galaxy surveys. The numerical implementation developed in this work is publicly released as the open-source Python package \texttt{ulptkit} (\url{https://github.com/naonori/ulptkit}).

\end{abstract}

%\keywords{Suggested keywords}%Use showkeys class option if keyword
                              %display desired
\maketitle

\section{Introduction}
\label{sec:intro}

The large-scale structure (LSS) of the Universe provides a powerful observational window into the physics of cosmic expansion and structure formation. Among various statistical observables, the matter power spectrum plays a central role by connecting the primordial fluctuations of the early Universe to the present-day distribution of matter.

To model the nonlinear evolution of the matter power spectrum, standard perturbation theory (SPT) has been widely employed (for a review, see Ref.~\cite{Bernardeau:2001qr}). However, SPT suffers from poor convergence at small scales due to long-wavelength displacements and nonlinear mode-coupling effects.

Lagrangian perturbation theory (LPT) offers a physically motivated alternative by treating the displacement vector as the primary dynamical variable. This perspective allows partial resummation of infrared (IR) contributions and has led to successful extensions such as the Zel'dovich approximation (ZA)~\cite{Zeldovich:1969sb}, convolution LPT (CLPT)~\cite{Carlson:2012bu} (see also~\cite{Sugiyama:2013mpa}), and IR-resummed approaches~\cite{Sugiyama:2013gza,Senatore:2014via,Baldauf:2015xfa,Blas:2016sfa,Senatore:2017pbn,Ivanov:2018gjr,Lewandowski:2018ywf,Sugiyama:2020uil}, all of which have demonstrated accurate predictions for the nonlinear damping of the baryon acoustic oscillation (BAO) feature~\cite{Sunyaev:1970eu,Peebles:1970ag}.

Despite recent developments, many existing LPT-based frameworks still face several limitations. While the conventional LPT provides a unified treatment of density fluctuations in real and redshift space, it does not offer a consistent formulation that incorporates post-reconstruction space on an equal footing. In particular, it has been pointed out that IR safety is violated in reconstructed fields~\cite{Sugiyama:2024eye}.

Moreover, when incorporating galaxy bias in LPT~\cite{Matsubara:2008wx}, one must account not only for the displacement field but also for biased density fluctuations defined in Lagrangian space. This additional ingredient partially breaks the structural unity of the LPT formulation.

Spectroscopic galaxy surveys measure the large-scale structure of the Universe in redshift space, where the observed galaxy density field is inevitably modulated by line-of-sight peculiar velocities. These distortions, known as redshift-space distortions (RSD)~\cite{Kaiser:1987qv}, introduce anisotropic features that must be accurately modeled to extract unbiased cosmological information from clustering measurements.

In parallel, density-field reconstruction techniques~\cite{Eisenstein:2006nk}, originally developed to sharpen the BAO feature by undoing large-scale bulk flows, have recently been shown to improve constraints on cosmological parameters derived from the broadband shape of the power spectrum~\cite{Hikage:2020fte,Wang:2022nlx}, beyond the BAO feature. Together, these observational advances underscore the need for a unified theoretical framework that consistently describes density fluctuations across real space, redshift space, and post-reconstruction space.

To address these challenges, we have recently developed \emph{Unified Lagrangian Perturbation Theory} (ULPT)~\cite{Sugiyama:inprep}, a new perturbative framework that enables a consistent and unified description of the density field in all three coordinate systems: real space, redshift space, and post-reconstruction space. Importantly, the ULPT formulation maintains IR safety across all three representations, ensuring that large-scale displacements are consistently resummed.

Unlike conventional approaches such as ZA, ULPT reformulates LPT by explicitly separating the density fluctuation into two physically distinct components: the \emph{Jacobian deviation}, which encapsulates intrinsic linear and nonlinear growth (including bias effects), and the \emph{displacement-mapping effect}, which accounts for large-scale convective distortions due to coordinate shifts.

This decomposition brings several theoretical advantages:
\begin{itemize}
    \item exact IR cancellation is realized nonperturbatively;
    \item the framework naturally recovers analytically derived IR-resummed models for BAO damping;
    \item the exponential damping observed in cross spectra between pre- and post-reconstruction fields emerges directly from the displacement-mapping term.
\end{itemize}

Furthermore, the ULPT formulation is inherently extensible. The effects of Lagrangian galaxy bias can be consistently encoded through additive operators in the Jacobian deviation. The structure of RSD is incorporated by modifying both the displacement vector and the Jacobian to account for velocity components along the line of sight. Finally, reconstruction enters the formalism as a shift applied solely to the displacement vector, leaving the Jacobian deviation unaffected. As such, ULPT provides a unified theoretical structure that allows bias, RSD, and reconstruction to be incorporated consistently within the same formalism.

While ULPT offers a theoretically robust foundation, its practical performance in modeling broadband power spectra, especially beyond the BAO scale, remains to be tested. This is of particular importance because broadband features also encode cosmological information beyond that captured by the BAO peak.

In this paper, we assess the predictive accuracy of ULPT by computing the one-loop matter power spectrum in real space and comparing it with both SPT and high-precision numerical predictions from the \textit{Dark Emulator}~\cite{Nishimichi:2018etk}. To further validate the robustness of our results, we perform an independent cross-check using the \textit{Euclid Emulator 2}~\cite{Euclid:2020rfv}, which has been carefully calibrated over a broad range of cosmological models.

In addition, we develop a fast and numerically stable algorithm for evaluating ULPT power spectra, based on efficient computation of the displacement-mapping integrals via Hankel transforms. This implementation provides a computationally efficient foundation for applying ULPT to future large-scale structure analyses.

Throughout this work, we adopt the fiducial cosmology implemented in the \textit{Dark Emulator} suite, which is consistent with the Planck 2015 best-fit $\Lambda$CDM model~\cite{Planck:2015fie}. The cosmological parameters are specified as follows: physical baryon density $\omega_b \equiv \Omega_b h^2 = 0.02225$, physical cold dark matter density $\omega_c \equiv \Omega_c h^2 = 0.1198$, dark energy density $\Omega_{\mathrm{de}} = 0.6844$, scalar spectral index $n_s = 0.9645$, amplitude of primordial curvature perturbations $\ln(10^{10} A_s) = 3.094$, dark energy equation-of-state parameter $w_0 = -1$, and total neutrino mass $\sum m_\nu = 0.06\,\mathrm{eV}$. The Hubble parameter is then determined to be $h = 0.6727$ from the flatness condition.

This paper is organized as follows. In Sec.~\ref{sec:ULPT}, we review the theoretical foundation of ULPT, emphasizing its decomposition into Jacobian deviation and displacement-mapping effects. Section~\ref{sec:approx_cov} presents a fast and numerically stable algorithm for evaluating the convolution integrals arising in the ULPT formulation. In Sec.~\ref{sec:sigma}, we compute the displacement correlation functions, which serve as key inputs to these integrals. Section~\ref{sec:FASTONE} describes the one-loop computation of the source power spectrum using a modified FAST-PT algorithm. In Sec.~\ref{sec:convergence}, we assess the convergence of the displacement-mapping expansion and determine an optimal truncation scheme. Section~\ref{sec:source_power} analyzes the relative roles of the source term and the displacement-mapping factor in shaping the full one-loop power spectrum. In Sec.~\ref{sec:emu}, we summarize the simulation-based emulators used for validation, along with their parameter coverage. Section~\ref{sec:BAO} focuses on the nonlinear damping of the BAO feature. The main results are presented in Sec.~\ref{sec:vs_emu}, where ULPT predictions are compared with emulator outputs across various cosmological models. Section~\ref{sec:future} outlines future directions, including extensions to biased tracers, RSD, and reconstruction, as well as two-loop improvements. We conclude in Sec.~\ref{sec:conclusion} with a brief summary.

\section{Unified Lagrangian Formalism}
\label{sec:ULPT}

In this section, we briefly review the core structure of ULPT~\cite{Sugiyama:inprep}. Throughout this paper, we omit the explicit time (or redshift) dependence of physical quantities for notational simplicity, unless stated otherwise.

\subsection{ULPT formulation of the Density Contrast}

Let $\rho(\boldsymbol{x})$ denote the dark matter density field at Eulerian position $\boldsymbol{x}$. The density contrast $\delta(\boldsymbol{x})$, defined with respect to the mean density $\bar{\rho}$, is given by
\begin{equation}
    \rho(\boldsymbol{x}) = \bar{\rho} \left[1 + \delta(\boldsymbol{x})\right].
    \label{eq:rho_delta}
\end{equation}

The mapping between Lagrangian coordinates $\boldsymbol{q}$ and Eulerian coordinates $\boldsymbol{x}$ is described by
\begin{equation}
    \boldsymbol{x} = \boldsymbol{q} + \boldsymbol{\Psi}(\boldsymbol{q}),
    \label{eq:x_q_map}
\end{equation}
where $\boldsymbol{\Psi}(\boldsymbol{q})$ is the Lagrangian displacement vector. The Jacobian relation between the volume elements is given by
\begin{equation}
    d^3x = J(\boldsymbol{q})\, d^3q, \qquad
    J(\boldsymbol{q}) \equiv \det\left(\frac{\partial \boldsymbol{x}}{\partial \boldsymbol{q}}  \right).
    \label{eq:jacobian}
\end{equation}

Mass conservation implies
\begin{equation}
    \rho(\boldsymbol{x})\, d^3x = \bar{\rho}\, d^3q.
    \label{eq:mass_conservation}
\end{equation}
From this, the density contrast evaluated at the Eulerian position $\boldsymbol{x} = \boldsymbol{q} + \boldsymbol{\Psi}(\boldsymbol{q})$ can be written as
\begin{equation}
    \delta(\boldsymbol{q} + \boldsymbol{\Psi}(\boldsymbol{q})) = \frac{1}{J(\boldsymbol{q})} - 1.
    \label{eq:delta_jacobian}
\end{equation}

To express the Eulerian density contrast $\delta(\boldsymbol{x})$ in terms of Lagrangian variables, we start from the identity
\begin{equation}
    \delta(\boldsymbol{x}) = \int d^3x' \, \delta(\boldsymbol{x}')\, \delta_{\rm D}(\boldsymbol{x} - \boldsymbol{x}'),
    \label{eq:identity}
\end{equation}
where $\delta_{\rm D}$ denotes the three-dimensional Dirac delta function. We then apply the coordinate transformation $\boldsymbol{x}' = \boldsymbol{q} + \boldsymbol{\Psi}(\boldsymbol{q})$, yielding
\begin{equation}
    \delta(\boldsymbol{x}) = \int d^3q\, \delta_{\rm J}(\boldsymbol{q})\, \delta_{\rm D}(\boldsymbol{x} - \boldsymbol{q} - \boldsymbol{\Psi}(\boldsymbol{q})),
    \label{eq:delta_realspace}
\end{equation}
where we have defined the \emph{Jacobian deviation} as
\begin{align}
    \delta_{\rm J}(\boldsymbol{q}) & \equiv J(\boldsymbol{q})\,\delta(\boldsymbol{q} + \boldsymbol{\Psi}(\boldsymbol{q}))
    \nonumber \\
    & = 1 - J(\boldsymbol{q}).
    \label{eq:jacobian_deviation}
\end{align}

Taking the Fourier transform of Eq.~\eqref{eq:delta_realspace}, we obtain the expression for the Fourier-space density contrast:
\begin{equation}
    \tilde{\delta}(\boldsymbol{k}) = \int d^3q\, e^{-i\boldsymbol{k} \cdot \boldsymbol{q}}\, e^{-i\boldsymbol{k} \cdot \boldsymbol{\Psi}(\boldsymbol{q})}\, \delta_{\rm J}(\boldsymbol{q}),
    \label{eq:delta_k}
\end{equation}
where we denote Fourier-transformed quantities with a tilde.

Equation~\eqref{eq:delta_k} provides the starting point for Unified Lagrangian Perturbation Theory (ULPT) in real space. In this formulation, the density contrast is explicitly decomposed into two physically distinct components:
\begin{itemize}
    \item \textbf{Jacobian deviation}: $\delta_{\rm J}(\boldsymbol{q})$ characterizes the intrinsic linear and nonlinear growth of structure.
    \item \textbf{Displacement-mapping effect}: the exponential factor $e^{-i\boldsymbol{k} \cdot \boldsymbol{\Psi}(\boldsymbol{q})}$ describes the nonlinear coordinate remapping of the Jacobian field induced by displacements fields.
\end{itemize}

This structural decomposition, which separates the density contrast into the Jacobian deviation and the displacement-mapping effect, can be extended to a variety of observational contexts beyond the case of dark matter in real space. Specifically:
\begin{itemize}
    \item \textbf{For biased tracers}, an additional contribution representing the biased fluctuation is linearly added to the Jacobian deviation, while the displacement-mapping effect retains the original structure.

    \item \textbf{For RSD}, both the Jacobian deviation and the displacement vector are modified to incorporate the effects of line-of-sight peculiar velocities. These modifications preserve the overall structure of the ULPT formulation while encoding the anisotropic nature of RSD.

    \item \textbf{For density-field reconstruction}, the displacement vector is modified by adding a reconstruction-induced shift field, whereas the Jacobian deviation is left unaltered. This structure ensures that the linear density field is preserved under reconstruction, as theoretically expected.
\end{itemize}

In this paper, we focus on the case of dark matter in real space for clarity.

\subsection{ULPT Formulation of the Power Spectrum}

The power spectrum in ULPT is computed as the Fourier transform of the two-point ensemble average of the density contrast. Starting from the expression
\begin{equation}
    P(\boldsymbol{k}) =
    \int d^3r \, e^{-i\boldsymbol{k}\cdot\boldsymbol{r}} \,
    \left\langle e^{-i\boldsymbol{k}\cdot\left[ \boldsymbol{\Psi}(\boldsymbol{q}) - \boldsymbol{\Psi}(\boldsymbol{q}') \right]}
    \delta_{\rm J}(\boldsymbol{q})
    \delta_{\rm J}(\boldsymbol{q}') \right\rangle,
\end{equation}
where $\boldsymbol{r} = \boldsymbol{q} - \boldsymbol{q}'$ is the separation vector between two Lagrangian positions, and $\langle \cdots \rangle$ denotes the ensemble average.

Defining
\begin{equation}
    X \equiv -i\boldsymbol{k} \cdot \left[ \boldsymbol{\Psi}(\boldsymbol{q}) - \boldsymbol{\Psi}(\boldsymbol{q}') \right],
\end{equation}
the power spectrum can be rewritten using cumulant expansion as
\begin{align}
    P(\boldsymbol{k}) &=
    \int d^3r \, e^{-i\boldsymbol{k}\cdot\boldsymbol{r}} \,
    \left\langle e^X \right\rangle \nonumber \\
    &\quad \times
    \left[
        \left\langle e^X \delta_{\rm J}(\boldsymbol{q}) \delta_{\rm J}(\boldsymbol{q}') \right\rangle_{\rm c}
        +
        \left\langle e^X \delta_{\rm J}(\boldsymbol{q}) \right\rangle_{\rm c}
        \left\langle e^X \delta_{\rm J}(\boldsymbol{q}') \right\rangle_{\rm c}
    \right],
\end{align}
where $\langle \cdots \rangle_{\rm c}$ denotes the connected part of the ensemble average.

The first factor, $\left\langle e^X \right\rangle$, depends only on the displacement field and is statistically uncorrelated with the Jacobian deviation field. We refer to this as the \emph{displacement-mapping factor}. Its cumulant expansion takes the form
\begin{align}
    \left\langle e^X \right\rangle = \exp\left[ -\overline{\Sigma}(\boldsymbol{k}) + \Sigma(\boldsymbol{k}, \boldsymbol{r}) \right],
\end{align}
where the exponent is defined as
\begin{align}
    -\overline{\Sigma}(\boldsymbol{k}) + \Sigma(\boldsymbol{k}, \boldsymbol{r})
    = \sum_{m=2}^{\infty} \frac{1}{m!}
    \left\langle
        \left[ -i\boldsymbol{k} \cdot (\boldsymbol{\Psi}(\boldsymbol{q}) - \boldsymbol{\Psi}(\boldsymbol{q}')) \right]^m
    \right\rangle_{\rm c}.
    \label{eq:Sigma_expand}
\end{align}
When considering the difference of a single displacement field, the displacement variance $\overline{\Sigma}(\boldsymbol{k})$ is defined as the zero-separation limit:
\begin{equation}
    \overline{\Sigma}(\boldsymbol{k}) \equiv \Sigma(\boldsymbol{k}, \boldsymbol{r} = 0).
\end{equation}

We define the remaining part of the integrand as the \emph{source correlation function}:
\begin{equation}
    \xi_{\rm J}(\boldsymbol{r}) \equiv
    \left\langle e^X \delta_{\rm J}(\boldsymbol{q}) \delta_{\rm J}(\boldsymbol{q}') \right\rangle_{\rm c}
    +
    \left\langle e^X \delta_{\rm J}(\boldsymbol{q}) \right\rangle_{\rm c}
    \left\langle e^X \delta_{\rm J}(\boldsymbol{q}') \right\rangle_{\rm c}.
\end{equation}

Expanding $e^X$ perturbatively, we obtain
\begin{align}
    \xi_{\rm J}(\boldsymbol{r}) &=
   \langle \delta_{\rm J}(\boldsymbol{q}) \delta_{\rm J}(\boldsymbol{q}') \rangle_{\rm c}
    +
    \langle (e^X - 1) \delta_{\rm J}(\boldsymbol{q}) \delta_{\rm J}(\boldsymbol{q}') \rangle_{\rm c} \nonumber \\
    &\quad +
    \langle (e^X - 1) \delta_{\rm J}(\boldsymbol{q}) \rangle_{\rm c}
    \langle (e^X - 1) \delta_{\rm J}(\boldsymbol{q}') \rangle_{\rm c},
    \label{eq:xi_J_def}
\end{align}
where we used $\langle \delta_{\rm J}\rangle=0$. This expression shows that the leading (i.e., linear-order) contribution to the source correlation function \(\xi_{\rm J}\) arises from the two-point correlation of the Jacobian deviation. Higher-order contributions to \(\xi_{\rm J}\) originate from two distinct sources: the nonlinear structure inherent in the Jacobian deviation's two-point correlation itself, and cross-correlations between the Jacobian deviation and the displacement field.

Combining the displacement-mapping factor and the source correlation function, the full power spectrum in ULPT is compactly expressed as
\begin{equation}
    P(\boldsymbol{k}) = e^{-\overline{\Sigma}(\boldsymbol{k})} \int d^3 r \, e^{-i \boldsymbol{k} \cdot \boldsymbol{r}} \, e^{\Sigma(\boldsymbol{k}, \boldsymbol{r})} \, \xi_{\rm J}(\boldsymbol{r}).
    \label{eq:Pk_ULPT}
\end{equation}
Although already introduced in the course of the derivation, the two physically distinct contributions to the ULPT power spectrum are restated here to highlight their conceptual significance:
\begin{itemize}
    \item \textbf{Source correlation function} $\xi_{\rm J}(r)$: the two-point correlation of the Jacobian deviation field, incorporating both linear and nonlinear intrinsic effects, as well as cross-correlations with the displacement field.

    \item \textbf{Displacement-mapping factor} $e^{ - \overline{\Sigma}(\boldsymbol{k})+\Sigma(\boldsymbol{k}, \boldsymbol{r})}$: an exponential modulation capturing the nonlinear influence of the displacement field on the clustering pattern. This factor isolates the convective coordinate remapping effect that is statistically uncorrelated with the intrinsic source field.
\end{itemize}
This structural separation plays a central role throughout the ULPT framework and enables a consistent treatment of IR effects.

We also define the \textit{source power spectrum} as the Fourier transform of the source correlation function:
\begin{equation}
    P_{\rm J}(\boldsymbol{k}) \equiv \int d^3 r \, e^{-i \boldsymbol{k} \cdot \boldsymbol{r}} \,
    \xi_{\rm J}(\boldsymbol{r}).
    \label{eq:source_power}
\end{equation}
With this definition, the total power spectrum can be systematically decomposed into the source power spectrum and an additional modulation term arising from the displacement-mapping factor acting on the source correlation function:
\begin{equation}
    P(\boldsymbol{k}) = P_{\rm J}(\boldsymbol{k}) + P_{\rm DM}(\boldsymbol{k}),
    \label{eq:P_J_DM}
\end{equation}
where the subscript ``DM'' stands for the ``displacement-mapping'' contribution. The explicit expression for $P_{\rm DM}(\boldsymbol{k})$ reads
\begin{align}
    P_{\rm DM}(\boldsymbol{k})  = 
    \int d^3 r \, e^{-i \boldsymbol{k} \cdot \boldsymbol{r}}
    \left[ e^{-\overline{\Sigma}(\boldsymbol{k})+ \Sigma(\boldsymbol{k}, \boldsymbol{r})} - 1 \right] 
    \xi_{\rm J}(\boldsymbol{r}).
    \label{eq:P_DM}
\end{align}

Alternatively, one may classify the contributions to the power spectrum based on whether they arise from convolution integrals associated with the displacement-mapping factor. In this case, the power spectrum is decomposed as
\begin{equation}
    P(\boldsymbol{k}) = P_{\rm CF}(\boldsymbol{k}) + P_{\rm CC}(\boldsymbol{k}),
    \label{eq:CF_CC}
\end{equation}
where ``CF'' denotes the ``convolution-free`` part and ''CC`` denotes the ''convolution-containing`` part. These are given explicitly by
\begin{align}
    P_{\rm CF}(\boldsymbol{k}) &= e^{-\overline{\Sigma}(\boldsymbol{k})} P_{\rm J}(\boldsymbol{k}), \nonumber \\
    P_{\rm CC}(\boldsymbol{k}) &= e^{-\overline{\Sigma}(\boldsymbol{k})}
    \int d^3 r \, e^{-i \boldsymbol{k} \cdot \boldsymbol{r}}
    \left[ e^{ \Sigma(\boldsymbol{k}, \boldsymbol{r})} - 1 \right] \xi_{\rm J}(\boldsymbol{r}).
    \label{eq:CFCC}
\end{align}

Since we are working in real space and assuming statistical isotropy, both $P(k)$ and $\xi_{\rm J}(r)$ depend only on the magnitudes of their respective arguments, $k = |\boldsymbol{k}|$ and $r = |\boldsymbol{r}|$. Furthermore, the displacement variance $\Sigma(\boldsymbol{k},\boldsymbol{r})$ depends only on three scalar quantities: $k$, $r$, and the scalar product $\hat{\boldsymbol{k}} \cdot \hat{\boldsymbol{r}}$. This property significantly simplifies the numerical evaluation of the convolution integrals, particularly when using spherical coordinates or Hankel transforms, as will be explicitly demonstrated in Section~\ref{sec:approx_cov}.

\section{Fast Algorithms for ULPT Convolution Integrals}
\label{sec:approx_cov}

We now present an efficient approximation scheme for evaluating the convolution integrals arising from the displacement-mapping factor in the ULPT formulation. This scheme closely follows the approach developed in Ref.~\cite{Sugiyama:2013mpa} for computing the ZA power spectrum as well as LPT power spectra including higher-order corrections, and recasts the relevant integrals in terms of Hankel transforms involving spherical Bessel functions.

Let \( f(r) \) be an arbitrary spherically symmetric function. The Hankel transform of order \( \ell \), commonly used in cosmology, is defined as
\begin{align}
    \tilde{f}_{\ell}(k) = (-i)^{\ell} (4\pi)  \int_0^\infty dr\, r^2\, j_{\ell}(kr)\, f(r),
\end{align}
where \( j_{\ell}(x) \) denotes the spherical Bessel function of order \( \ell \).
The corresponding inverse transform is given by
\begin{align}
    f(r) = i^{\ell} \int \frac{dk\, k^2}{2\pi^2} j_{\ell}(kr)\, \tilde{f}_{\ell}(k).
\end{align}
These transforms can be efficiently evaluated using FFTLog-based algorithms~\cite{Hamilton:1999uv}. In our implementation, we make use of the publicly available \texttt{mcfit} package~\cite{mcfit}.

The goal of this paper is to compute the ULPT power spectrum including one-loop corrections. Our approximation exactly reproduces the one-loop result of SPT, while resumming higher-order terms that originate from the convolution structure of the displacement-mapping factor.

To this end, we expand the exponent of the displacement-mapping factor at linear order and compute the source correlation function up to one-loop order:
\begin{align}
    \Sigma(\boldsymbol{k},\boldsymbol{r}) & = \Sigma^{(\mathrm{lin})}(\boldsymbol{k},\boldsymbol{r}), \\
    \xi_{\mathrm{J}}(r) & = \xi_{\rm J}^{(\mathrm{lin})}(r) + \xi^{(1\text{-loop})}_{\mathrm{J}}(r).
\end{align}
At linear order, the source correlation function $\xi_{\rm J}^{(\mathrm{lin})}(r)$ coincides with the standard matter correlation function, and we denote it simply by $\xi_{\mathrm{lin}}(r)$ in the following.

The corresponding source power spectrum reads
\begin{align}
    P_{\mathrm{J}}(k) = P_{\mathrm{lin}}(k) + P^{(1\text{-loop})}_{\mathrm{J}}(k),
\end{align}
where $P_{\rm lin}(k)$ is the linear dark matter power spectrum.

The linear-order displacement variance is given by
\begin{equation}
    \Sigma^{(\mathrm{lin})}(\boldsymbol{k}, \boldsymbol{r})
    = \int \frac{d^3k'}{(2\pi)^3} e^{i\boldsymbol{k}'\cdot\boldsymbol{r}}
    \left( \frac{\boldsymbol{k}\cdot\boldsymbol{k}'}{k'^2} \right)^2\, P_{\rm lin}(k').
    \label{eq:Sigma_lin}
\end{equation}
Performing the angular integrals analytically, it becomes
\begin{equation}
    \Sigma^{(\mathrm{lin})}(\boldsymbol{k}, \boldsymbol{r}) = k^2\left[ \sigma_0^2(r) + 2 \mathcal{L}_2(\hat{\boldsymbol{k}} \cdot \hat{\boldsymbol{r}}) \sigma_2^2(r) \right],
    \label{eq:Sigma_lin_ana}
\end{equation}
with
\begin{align}
    \sigma_{\ell}^2(r) = \frac{1}{3} i^{\ell} \int \frac{dk}{2\pi^2} j_{\ell}(kr) P_{\mathrm{lin}}(k).
    \label{eq:sigma_l}
\end{align}
Here, $\mathcal{L}_{\ell}$ is the Legendre polynomial of order $\ell$. The variance $\overline{\Sigma}(k)$ is defined via the zero-separation limit:
\begin{align}
    \bar{\sigma}^2 = \sigma_0^2(0) = \frac{1}{3} \int \frac{dk}{2\pi^2} P_{\mathrm{lin}}(k),
    \label{eq:sigma_bar}
\end{align}
leading to
\begin{align}
    \overline{\Sigma}^{(\rm lin)}(k) = k^2 \bar{\sigma}^2.
\end{align}

Substituting these results into the general expression for the ULPT power spectrum, we obtain:
\begin{align}
    P(k) = e^{-k^2 \bar{\sigma}^2} \int d^3 r\, e^{-i\boldsymbol{k} \cdot \boldsymbol{r}}\,
    e^{k^2[ \sigma_0^2(r) + 2 \mathcal{L}_2(\mu) \sigma_2^2(r) ]} \xi_{\mathrm{J}}(r),
    \label{eq:P_ULPT_1loop}
\end{align}
where $\mu=\hat{\boldsymbol{k}}\cdot\hat{\boldsymbol{r}}$.

We now decompose the total power spectrum into the CF and CC contributions, as defined in Eq.~\eqref{eq:CFCC}:
\begin{align}
    P_{\mathrm{CF}}(k) &= e^{-k^2 \bar{\sigma}^2} P_{\mathrm{J}}(k), \nonumber \\
    P_{\mathrm{CC}}(k) &= e^{-k^2 \bar{\sigma}^2} \int d^3 r\, e^{-i\boldsymbol{k} \cdot \boldsymbol{r}} \nonumber \\
    &\quad \times \left[ e^{k^2[ \sigma_0^2(r) + 2 \mathcal{L}_2(\mu) \sigma_2^2(r) ]} - 1 \right] \xi_{\mathrm{J}}(r).
\end{align}

To further simplify the CC term, we expand the angular dependence in the exponent:
\begin{align}
    e^{2k^2 \mathcal{L}_2(\mu) \sigma_2^2(r)} = \sum_{n=0}^\infty \frac{1}{n!} [2k^2 \mathcal{L}_2(\mu) \sigma_2^2(r)]^n,
\end{align}
which leads to the decomposition
\begin{align}
    P_{\mathrm{CC}}(k) = \sum_{n=0}^\infty P_{\mathrm{CC}}^{[n]}(k),
\end{align}
with the zeroth and higher-order terms given by
\begin{align}
    P_{\mathrm{CC}}^{[0]}(k) &= e^{-k^2 \bar{\sigma}^2} \int d^3 r\, e^{-i \boldsymbol{k} \cdot \boldsymbol{r}}
    \left[ e^{k^2 \sigma_0^2(r)} - 1 \right] \xi_{\mathrm{J}}(r), \nonumber \\
    P_{\mathrm{CC}}^{[n\geq1]}(k) &= e^{-k^2 \bar{\sigma}^2} \int d^3 r\, e^{-i \boldsymbol{k} \cdot \boldsymbol{r}} e^{k^2 \sigma_0^2(r)} \nonumber \\
    &\quad \times \frac{1}{n!} [2k^2 \mathcal{L}_2(\mu) \sigma_2^2(r)]^n \xi_{\mathrm{J}}(r).
\end{align}

Performing the angular integrals analytically, we recast these expressions as Hankel transforms:
\begin{align}
    P_{\mathrm{CC}}^{[0]}(k) &= e^{-k^2 \bar{\sigma}^2} (4\pi) \int dr\, r^2 \left[ e^{k^2 \sigma_0^2(r)} - 1 \right] j_0(kr) \xi_{\mathrm{J}}(r), \nonumber \\
    P_{\mathrm{CC}}^{[n\geq1]}(k) &= \frac{1}{n!} (2k^2)^n e^{-k^2 \bar{\sigma}^2} (4\pi) \int dr\, r^2 e^{k^2 \sigma_0^2(r)} \nonumber \\
    &\quad \times I^{[n]}(k, r) \left[ \sigma_2^2(r) \right]^n \xi_{\mathrm{J}}(r).
\end{align}

The angular integrals $I^{[n]}(k, r)$ are defined by
\begin{align}
    I^{[n]}(k, r) = \int_{-1}^1 \frac{d\mu}{2} e^{-ikr\mu} \left[ \mathcal{L}_2(\mu) \right]^n,
\end{align}
and for $n = 1$ to $4$ they evaluate to
\begin{align}
    I^{[1]}(k, r) &= -j_2(kr), \nonumber \\
    I^{[2]}(k, r) &= \frac{1}{5} j_0(kr) - \frac{2}{7} j_2(kr) + \frac{18}{35} j_4(kr),  \nonumber \\
    I^{[3]}(k, r) &= \frac{2}{35} j_0(kr) - \frac{291}{154} j_2(kr) \nonumber \\
    & \quad + \frac{756}{2695} j_4(kr) - \frac{18}{77} j_6(kr),  \nonumber \\
    I^{[4]}(k, r) &= \frac{3}{35} j_0(kr) - \frac{133}{512} j_2(kr) + \frac{687}{1872} j_4(kr) \nonumber \\
    &\quad - \frac{1441}{7704} j_6(kr) + \frac{1451}{14408} j_8(kr).
\end{align}

These results demonstrate that the convolution contributions can be systematically evaluated using Hankel transforms.

\section{Correlation Functions of the Displacement Field}
\label{sec:sigma}

\begin{figure}[t]
    \centering
    \includegraphics[width=\columnwidth]{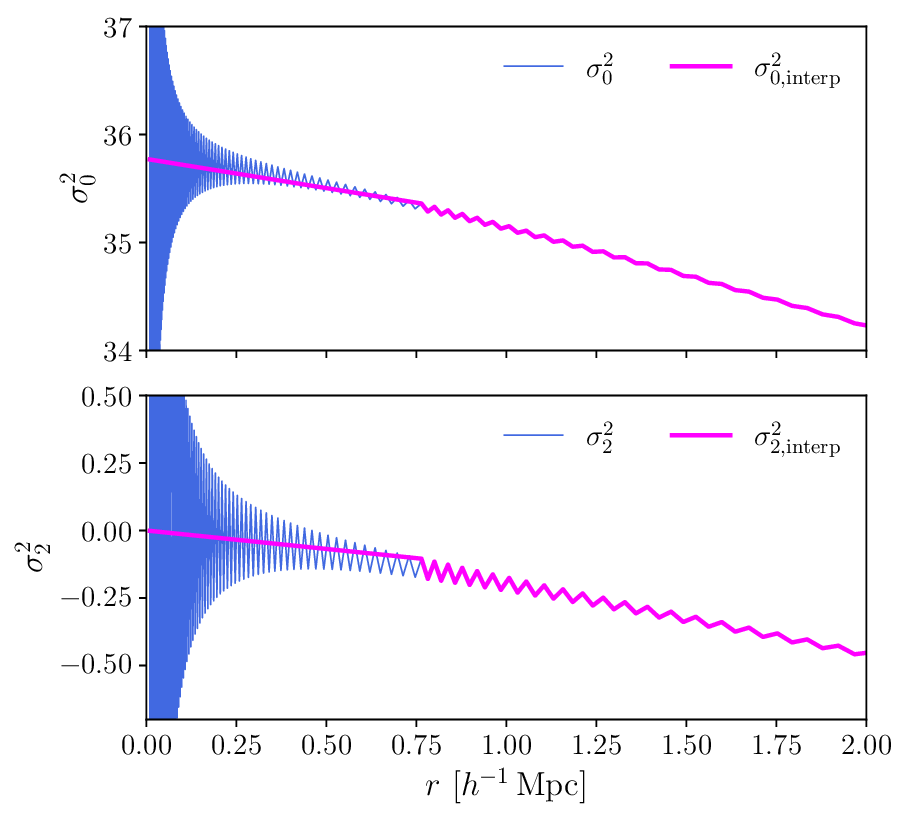}
    \caption{
        Displacement correlation functions \( \sigma_0^2(r) \) and \( \sigma_2^2(r) \), computed at redshift \( z = 0 \) from the linear power spectrum using FFTLog. This figure focuses on the small-scale regime, $r \leq 2\,h^{-1}\mathrm{Mpc}$. The original FFTLog results, which exhibit spurious oscillations at small separations due to the finite \( k_{\rm max} \) cutoff, are shown as blue curves. To suppress these instabilities, we apply a linear interpolation between \( r = 0.75\,h^{-1}\mathrm{Mpc} \) and \( r = 0 \), connecting the analytic values at the origin to the numerical ones at the cutoff scale.
}
    \label{fig:sigma_small}
\end{figure}

In this section, we present the numerical evaluation of the displacement correlation functions \( \sigma^2_\ell(r) \) for \( \ell = 0 \) and \( \ell = 2 \), as defined in Eq.~\eqref{eq:sigma_l}. These quantities characterize the scale-dependent contributions of the displacement field and play a central role in evaluating the convolution integral induced by the displacement-mapping factor in ULPT.

The results shown here correspond to redshift \( z = 0 \), where nonlinear effects are most pronounced. The correlation functions are efficiently computed using the FFTLog algorithm~\cite{Hamilton:1999uv,mcfit}. However, at small separations \( r \), the finite \( k_{\rm max} \) cutoff in the input linear power spectrum induces spurious oscillations in the numerical results. To suppress such artifacts, we adopt a conservative value of \( k_{\rm max} = 100\,h\,\mathrm{Mpc}^{-1} \), which stabilizes the evaluation down to sub-megaparsec scales.

Figure~\ref{fig:sigma_small} shows that numerical instabilities become significant for \( r \lesssim 0.75\,h^{-1}\mathrm{Mpc} \), where both \( \sigma_0^2(r) \) and \( \sigma_2^2(r) \) exhibit artificial oscillations. Nevertheless, the limiting values at \( r = 0 \) are known analytically: \( \sigma_0^2(0) = \bar{\sigma}^2 \), as given in Eq.~\eqref{eq:sigma_bar}, and \( \sigma_2^2(0) = 0 \). We therefore apply a linear interpolation between \( r = 0 \) and \( r = 0.75\,h^{-1}\mathrm{Mpc} \) to smoothly connect the stable analytic values at the origin to the numerically computed values at the cutoff scale. This procedure effectively eliminates the unphysical FFT-induced fluctuations in the small-\( r \) regime and ensures numerical stability in subsequent integrations.

\begin{figure}[t]
    \centering
    \includegraphics[width=\columnwidth]{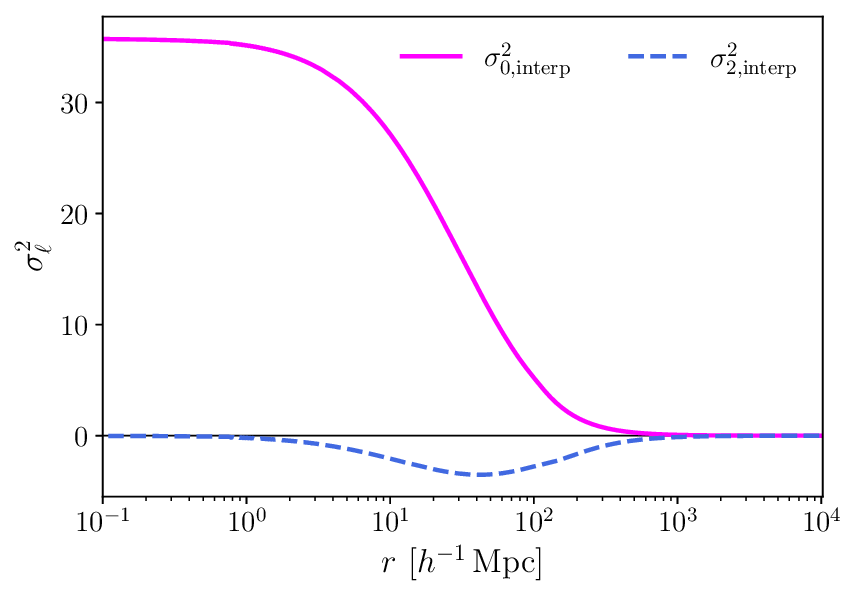}
    \caption{Full-range displacement correlation functions at $z=0$.
        The magenta solid line shows the monopole component $\sigma_0^{2}(r)$,
        and the blue dashed line shows the quadrupole component $\sigma_2^{2}(r)$.
        The small-scale behavior for $r < 0.75\,h^{-1}\mathrm{Mpc}$ is smoothly
        interpolated as described in Fig.~\ref{fig:sigma_small}.
}
    \label{fig:sigma_large}
\end{figure}

Figure~\ref{fig:sigma_large} shows the displacement correlation functions $\sigma_\ell^2(r)$ over a wide range of comoving separations, spanning from $r = 0.1\,h^{-1}\mathrm{Mpc}$ to $10^4\,h^{-1}\mathrm{Mpc}$. For $r < 0.75\,h^{-1}\mathrm{Mpc}$, the curves are linearly interpolated between the numerical values at $r = 0.75\,h^{-1}\mathrm{Mpc}$ and the analytic limits at $r = 0$, following the same procedure used in Fig.~\ref{fig:sigma_small}.

As required by definition, the monopole component $\sigma_0^2(r)$ asymptotically approaches $\bar{\sigma}^2$ at small separations, but rapidly decreases between $r \sim 10\,h^{-1}\mathrm{Mpc}$ and $r \sim 100\,h^{-1}\mathrm{Mpc}$, eventually approaching zero toward large scales. In contrast, the quadrupole component $\sigma_2^2(r)$ exhibits a broad negative trough around $r \sim 50\,h^{-1}\mathrm{Mpc}$, and vanishes both in the small- and large-scale limits.

When comparing the relative amplitudes of the monopole and quadrupole components, it is important to note that the monopole contribution always appears in the combination $\sigma_0^2(r) - \bar{\sigma}^2$ in the convolution integrals. Therefore, the relevant quantity to assess the hierarchy is the ratio $|\sigma_2^2(r)/[\sigma_0^2(r) - \bar{\sigma}^2]|$. This ratio remains below $\sim 0.2$ for $r \gtrsim 10\,h^{-1}\mathrm{Mpc}$, indicating that the quadrupole contribution is significantly suppressed relative to the effective monopole term over the scales of interest. This provides numerical support for the perturbative expansion scheme adopted in Sec.~\ref{sec:approx_cov}, in which the $\sigma_2^2$-dependent terms are systematically expanded.

\section{Fast Evaluation of the One-Loop Source Power Spectrum}
\label{sec:FASTONE}

In this section, we develop a fast and numerically stable algorithm for computing the one-loop source power spectrum $P_{\rm J}$ within the ULPT framework. We begin with a brief overview of the publicly available \texttt{FAST-PT} algorithm~\cite{McEwen:2016fjn,Fang:2016wcf} in Sec.~\ref{sec:FAST_PT}, and in Sec.~\ref{sec:source_P} we describe how the same algorithm can be applied to ULPT by replacing the standard SPT kernel with its ULPT counterpart.

\subsection{Review of the FAST-PT Algorithm}
\label{sec:FAST_PT}

\texttt{FAST-PT}~\cite{McEwen:2016fjn} is a Python-based tool designed to efficiently compute convolution integrals which arise in cosmological perturbation theory. The method is particularly well-suited for applications involving the decomposition of angular dependencies in mode-coupling kernels into Legendre polynomials, enabling analytic evaluation of angular integrals. The remaining radial integrals, expressed as Hankel transforms involving spherical Bessel functions, can be rapidly evaluated using FFTLog techniques. Consequently, the overall computational cost is reduced to $\mathcal{O}(N \log N)$.

For the one-loop matter power spectrum in SPT, the total correction is decomposed as
\begin{equation}
    P^{(\text{1-loop})}(k) = P^{(22)}(k) + P^{(13)}(k),
    \label{eq:P_22_13}
\end{equation}
where $P^{(22)}$ represents a double convolution of the linear power spectrum and $P^{(13)}$ captures the coupling between linear and third-order terms. In FAST-PT, these terms are regularized to avoid IR divergences:
\begin{align}
    P^{(22)}_{\text{reg}}(k) &= P^{(22)}(k) - k^2 \bar{\sigma}^2 P_{\text{lin}}(k), \\
    P^{(13)}_{\text{reg}}(k) &= P^{(13)}(k) + k^2 \bar{\sigma}^2 P_{\text{lin}}(k),
\end{align}
where $\bar{\sigma}^2$ is the displacement variance given by Eq.~\eqref{eq:sigma_bar}. This procedure preserves the total one-loop correction while canceling unphysical IR contributions in each term.

For the $P^{(13)}_{\text{reg}}$ term, the regularized expression takes the form
\begin{equation}
    P_{\text{reg}}^{(13)}(k) = 
    \frac{k^3}{252(2\pi)^2} P_{\mathrm{lin}}(k) \int_0^\infty dr\, r^2\, P_{\mathrm{lin}}(kr)\, Z_{\mathrm{reg}}(r),
\end{equation}
where the kernel $Z_{\mathrm{reg}}(r)$ is given by
\begin{align}
    Z_{\mathrm{reg}}(r) &= \frac{12}{r^4} + \frac{10}{r^2} + 100 - 42r^2 \nonumber \\
    &\quad + \frac{3}{r^5}(7r^2 + 2)(r^2 - 1)^3 \ln\left|\frac{r+1}{r-1}\right|.
\end{align}
Although this integral naively appears to scale as $\mathcal{O}(N^2)$, it can be rewritten as a discrete convolution using logarithmic variable transformations and then evaluated in $\mathcal{O}(N \log N)$ operations using FFT techniques.

The $P^{(22)}$ term poses additional challenges due to its dependence on two distinct wavevectors. FAST-PT circumvents this by transforming the convolution to configuration space using Legendre expansions. In configuration space, $P^{(22)}$ is evaluated as
\begin{equation}
    P^{(22)}(k) = 4\pi \int dr\, r^2\, j_0(kr)\, \xi^{(22)}(r),
\end{equation}
where
\begin{align}
    \xi^{(22)}(r) &= 2 \int \frac{d^3k_1}{(2\pi)^3} \int \frac{d^3k_2}{(2\pi)^3} 
    e^{i\boldsymbol{k}_1\cdot\boldsymbol{r}} e^{i\boldsymbol{k}_2\cdot\boldsymbol{r}} \nonumber \\
    &\quad \times \left[ F_2(\boldsymbol{k}_1, \boldsymbol{k}_2) \right]^2 
    P_{\mathrm{lin}}(k_1) P_{\mathrm{lin}}(k_2),
\end{align}
and the second-order kernel is
\begin{equation}
    F_2(\boldsymbol{k}_1, \boldsymbol{k}_2) = \frac{5}{7} + \frac{1}{2}\mu_{12} \left( \frac{k_1}{k_2}+\frac{k_2}{k_1} \right) + \frac{2}{7} \mu_{12}^2,
    \label{eq:F2}
\end{equation}
with $\mu_{12} = \hat{\boldsymbol{k}}_1 \cdot \hat{\boldsymbol{k}}_2$ denoting the cosine of the angle between the wavevectors. 

The kernel squared can be expanded in Legendre polynomials:
\begin{equation}
    \left[ F_2(\boldsymbol{k}_1,\boldsymbol{k}_2) \right]^2 = \sum_{\alpha,\beta,\ell} c_{\alpha\beta\ell}\, k_1^{\alpha} k_2^{\beta} \mathcal{L}_{\ell}(\mu_{12}),
\end{equation}
where the coefficients $c_{\alpha\beta\ell}$ are listed in Table~\ref{tab:c_alpha_beta_l}. This expansion allows the angular integrations to be performed analytically. The resulting correlation function is then written as
\begin{equation}
    \xi^{(22)}(r) = \sum_{\alpha,\beta,\ell} 2\,c_{\alpha\beta\ell}\, J_{\alpha\beta\ell}(r),
\end{equation}
with each term defined by
\begin{align}
J_{\alpha\beta\ell}(r) &= 
\left[ i^{\ell}\int \frac{dk_1\, k_1^2}{2\pi^2}\, k_1^{\alpha} P_{\mathrm{lin}}(k_1)\, j_\ell(k_1 r) \right] \nonumber \\
&\quad \times
\left[ i^{\ell}\int \frac{dk_2\, k_2^2}{2\pi^2}\, k_2^{\beta} P_{\mathrm{lin}}(k_2)\, j_\ell(k_2 r) \right].
\label{eq:J_abl}
\end{align}
These integrals are efficiently evaluated using Hankel transforms, leading to substantial computational speedup compared to brute-force multidimensional integration.

Among these terms, the component with $(\alpha,\beta,\ell) = (2, -2, 0)$ is known to be IR-sensitive. To address this, FAST-PT applies a regularization procedure by analytically subtracting the divergent component at the integrand level. The regularized form of $J_{2,-2,0}(r)$ is given by
\begin{align}
    J^{\rm reg}_{2,-2,0}(r) &= 
\left[ \int \frac{dk_1\, k_1^2}{2\pi^2}\, k_1^{2} P_{\mathrm{lin}}(k_1)\, j_0(k_1 r) \right] \nonumber \\
&\quad \times\left[ \int \frac{dk_2}{2\pi^2} P_{\mathrm{lin}}(k_2)\, \left\{j_0(k_2 r) - 1 \right\}\right].
\end{align}
where the subtraction of unity ensures convergence in the IR limit, since $[j_0(k_2 r) - 1] = (k_2 r)^2/6 + \cdots$ for small $k_2$.

The regularization terms such as $k^2 \bar{\sigma}^2 P_{\mathrm{lin}}(k)$ correspond to the effect of long-wavelength bulk flows. These displacements induce apparent power at large scales, which must cancel in the final, physically meaningful power spectrum due to statistical homogeneity and Galilean invariance. While the total one-loop power spectrum does exhibit this cancellation, individual components such as $P_{13}$ and $P_{22}$ do not, and must therefore be regularized explicitly.

From a numerical standpoint, even though the linear matter power spectrum in a $\Lambda$CDM cosmology behaves as $P_{\mathrm{lin}}(k) \propto k^{n_s}$ with $n_s \simeq 0.96$ in the $k \to 0$ limit, and thus poses no true divergence, the FFT-based approach used in FAST-PT introduces a logarithmic discretization. This results in a log-periodic effective power spectrum with artificial power at extremely large or small scales. Consequently, these spurious contributions must be analytically subtracted before integration to avoid numerical instabilities.

\renewcommand{\arraystretch}{1.6}
\begin{table}[!htbp]
    \centering
    \caption{
        Coefficients $c_{\alpha\beta\ell}$ and $c_{{\rm J},\alpha\beta\ell}$ used in the Legendre expansion of mode-coupling kernels relevant to the 22-type term in the one-loop power spectrum. 
The coefficients $c_{\alpha\beta\ell}$ appear in the standard SPT expression for the one-loop power spectrum, while $c_{ {\rm J}, \alpha\beta\ell}$ correspond to the contributions retained in the ULPT source term. In particular, the IR-sensitive component $(\alpha,\beta,\ell) = (2, -2, 0)$ is excluded in the ULPT formulation, leading to automatic IR regularization.
}

\label{tab:c_alpha_beta_l}
    \begin{tabular}{ccc|cc}
        \hline
        $\alpha$ & $\beta$ & $\ell$ & $c_{\alpha\beta\ell}$ & $c_{ {\rm J}, \alpha\beta\ell}$\\
        \hline
        0 & 0 & 0  &  $\frac{1219}{1470}$ & $\frac{242}{735}$\\
        0 & 0 & 2  &  $\frac{671}{1029}$  & $\frac{671}{1029}$\\
        0 & 0 & 4  &  $\frac{32}{1715}$   & $\frac{32}{1715}$\\
        1 & -1 & 1 & $\frac{62}{35}$  & $\frac{27}{35}$\\
        1 & -1 & 3 & $\frac{8}{35}$  & $\frac{8}{35}$\\
        2 & -2 & 0 & $\frac{1}{6}$  & $0$ \\
        2 & -2 & 2 & $\frac{1}{3}$  & $0$\\
        \hline
    \end{tabular}
\end{table}

\subsection{FAST-PT Implementation with ULPT-Specific Kernels}
\label{sec:source_P}

In our implementation, we retain the core algorithmic structure of \texttt{FAST-PT} and adapt it to the ULPT framework by substituting the standard $F_2$ kernel with the ULPT-specific version. No modification to the underlying numerical techniques is required.

In the ULPT formulation, the SPT one-loop power spectrum can be decomposed, following Eq.~\eqref{eq:P_J_DM}, into two distinct contributions:
\begin{equation}
    P^{(1\text{-loop})}(k) = P_{\rm J}^{(1\text{-loop})}(k)
    + P_{\rm DM}^{(1\text{-loop})}(k),
    \label{eq:SPT_J_DM}
\end{equation}
where the displacement-mapping (DM) contribution is given by
\begin{align}
    P_{\rm DM}^{(1\text{-loop})}(k)
    &= \int d^3 r\,  e^{-i\boldsymbol{k}\cdot\boldsymbol{r}} \nonumber \\
    &\quad \times
    \left[ -\overline{\Sigma}^{\rm (lin)}(\boldsymbol{k}) + \Sigma^{\rm (lin)}(\boldsymbol{k},\boldsymbol{r}) \right]
    \xi_{\rm lin}(r).
    \label{eq:P_DM_1loop}
\end{align}
Accordingly, it is not necessary to directly evaluate Eq.~\eqref{eq:xi_J_def}, which defines the source correlation function. Instead, the one-loop source power spectrum can be obtained by subtracting the displacement-mapping contribution from the known one-loop result in SPT:%
\footnote{At two-loop order, the source power spectrum is given by
\begin{equation}
P_{\rm J}^{(2\text{-loop})}(k) = P^{(2\text{-loop})}(k) - P_{\rm DM}^{(2\text{-loop})}(k), \nonumber
\end{equation}
where the two-loop DM contribution can be expressed in terms of the linear and one-loop results as
\begin{align}
P_{\rm DM}^{(2\text{-loop})}(k)
&= \int d^3 r\, e^{-i\boldsymbol{k}\cdot\boldsymbol{r}} \Bigg\{ \left[ -\overline{\Sigma}^{\rm (lin)}(\boldsymbol{k}) + \Sigma^{\rm (lin)}(\boldsymbol{k},\boldsymbol{r}) \right] \xi_{\rm J}^{(1\text{-loop})}(r) \nonumber \\
&\quad + \frac{1}{2} \left[ -\overline{\Sigma}^{\rm (lin)}(\boldsymbol{k}) + \Sigma^{\rm (lin)}(\boldsymbol{k},\boldsymbol{r}) \right]^2 \xi_{\rm lin}(r) \nonumber \\
&\quad + \left[ -\overline{\Sigma}^{(1\text{-loop})}(\boldsymbol{k}) + \Sigma^{(1\text{-loop})}(\boldsymbol{k},\boldsymbol{r}) \right] \xi_{\rm lin}(r) \Bigg\}. \nonumber
\label{eq:P_DM_2loop}
\end{align}
}
\begin{equation}
P_{\rm J}^{(1\text{-loop})}(k) = P^{(1\text{-loop})}(k) - P_{\rm DM}^{(1\text{-loop})}(k).
\end{equation}

As in the SPT framework in Eq.~\eqref{eq:P_22_13}, Eq.~\eqref{eq:P_DM_1loop} can be decomposed into two components: the so-called 13-type and 22-type terms, corresponding to the contributions without and with mode-coupling integrals, respectively. In the ULPT formulation, the term involving $-\overline{\Sigma}^{\rm (lin)}(\boldsymbol{k})$ corresponds to the 13-type contribution, while the term involving $\Sigma^{\rm (lin)}(\boldsymbol{k},\boldsymbol{r})$ corresponds to the 22-type contribution.

Using Eqs.~\eqref{eq:Sigma_lin} and \eqref{eq:sigma_bar}, the individual DM terms can be expressed as:
\begin{align}
    P_{\rm DM}^{(13)}(k) &= -k^2\bar{\sigma}^2 P_{\rm lin}(k), \nonumber \\
    P_{\rm DM}^{(22)}(k) &= 4\pi \int dr\, r^2\, j_0(kr)\, \xi_{\rm DM}^{(22)}(r),
    \label{eq:PDM13DM22}
\end{align}
where
\begin{align}
    \xi_{\rm DM}^{(22)}(r) &= \int \frac{d^3k_1}{(2\pi)^3} \int \frac{d^3k_2}{(2\pi)^3} 
    e^{i\boldsymbol{k}_1\cdot\boldsymbol{r}} e^{i\boldsymbol{k}_2\cdot\boldsymbol{r}} \nonumber \\
    &\quad \times \left( \frac{\boldsymbol{k}\cdot\boldsymbol{k}_2}{k_2^2} \right)^2
    P_{\mathrm{lin}}(k_1) P_{\mathrm{lin}}(k_2).
    \label{eq:xi_DM_22}
\end{align}

Accordingly, the 13-type and 22-type components of the source power spectrum can be defined by subtracting the DM contributions from the SPT terms:
\begin{equation}
    P_{\rm J}^{(\text{1-loop})}(k) = P_{\rm J}^{(13)}(k) + P_{\rm J}^{(22)}(k) ,
    \label{eq:1loop_J}
\end{equation}
where
\begin{align}
    P_{\rm J}^{(13)}(k) &= P^{(13)}(k) - P_{\rm DM}^{(13)}(k), \nonumber \\
    P_{\rm J}^{(22)}(k) &= P^{(22)}(k) - P_{\rm DM}^{(22)}(k).
    \label{eq:P_J13_P_J22}
\end{align}

It is important to note that the source 13-type term, $P_{\rm J}^{(13)}$, coincides with the regularized version of the SPT 13-type contribution as defined in \texttt{FAST-PT}:
\begin{equation}
    P_{\rm J}^{(13)}(k) = P_{\rm reg}^{(13)}(k).
\end{equation}
Therefore, for this term, we can directly adopt the result provided by \texttt{FAST-PT}.

The 22-type component of the DM contribution, $P_{\rm DM}^{(22)}$, is computed in configuration space following the methods outlined in Sec.~\ref{sec:FAST_PT}. Its inverse Fourier transform, given in Eq.~\eqref{eq:xi_DM_22}, is decomposed in terms of Legendre polynomials as follows:
\begin{align}
    \xi_{\rm DM}^{(22)}(r) & = 2\Bigg[ \frac{1}{2} J_{0,0,0}(r)
    + J_{1,-1,1}(r) \nonumber \\
& \quad + \frac{1}{3}J_{2,-2,2}(r) + \frac{1}{6} J_{2,-2,0}(r) \Bigg],
\end{align}
where the functions $J_{\alpha\beta\ell}(r)$ are defined in Eq.~\eqref{eq:J_abl}. Substituting this result into Eq.~\eqref{eq:P_J13_P_J22}, the inverse Fourier transform of the source 22-type component becomes
\begin{equation}
    \xi_{\rm J}^{(22)}(r) = \sum_{\alpha,\beta,\ell} 2\, c_{{\rm J},\alpha\beta\ell} J_{\alpha\beta\ell}(r),
\end{equation}
where the coefficients $c_{{\rm J},\alpha\beta\ell}$ are listed in Table~\ref{tab:c_alpha_beta_l}.

Notably, in $P_{\rm J}^{(22)}$, the IR-sensitive term $J_{2,-2,0}(r)$ appears with a zero coefficient. As a result, the source power spectrum $P_{\rm J}^{(22)}$ is automatically regularized.

This natural regularization arises from the fact that IR-sensitive contributions, such as those due to long-wavelength bulk flows, are fully captured by the displacement-mapping factor. These effects are thus separated from the source power spectrum and handled analytically.

In particular, numerical instabilities in the IR regime are effectively absorbed into the displacement-mapping exponent, $\sigma_0^2(r)$ in Eq.~\eqref{eq:sigma_l}. However, because $\sigma_0^2(r)$ always appears in the combination $\sigma_0^2(r) - \bar{\sigma}^2$, the integrands involve the form $[j_0(kr) - 1]$, ensuring convergence in the IR limit.

As shown in Sec.~\ref{sec:sigma}, in this work, the function $\sigma_0^2(r)$ is computed directly using FFTLog, implemented via the \texttt{mcfit} package~\cite{mcfit}. The zero-separation variance $\bar{\sigma}^2$ is evaluated independently. As a result, we do not apply the analytic regularization procedure described above, which may lead to numerical instability at small values of $r$. To mitigate this issue, we extend the linear input power spectrum up to a maximum wavenumber of $k_{\rm max} = 100\,h\,{\rm Mpc}^{-1}$. Furthermore, for $r < 0.75\,h^{-1}{\rm Mpc}$, we linearly interpolate $\sigma_0^2(r)$ between its value at $r = 0.75\,h^{-1}{\rm Mpc}$ and the limiting value $\bar{\sigma}^2$ at $r = 0$. Similar numerical instabilities are also observed in $\sigma_2^2(r)$, and we apply the same interpolation procedure to this quantity as well. Since our final results target scales up to $k \simeq 0.4\,h\,{\rm Mpc}^{-1}$, we expect that contributions from $r < 0.75\,h^{-1}{\rm Mpc}$ do not significantly affect the accuracy. 

\section{Convergence of the Expansion Scheme for Displacement-Mapping Integrals}
\label{sec:convergence}

\begin{figure}[!t]
    \centering
    \includegraphics[width=\columnwidth]{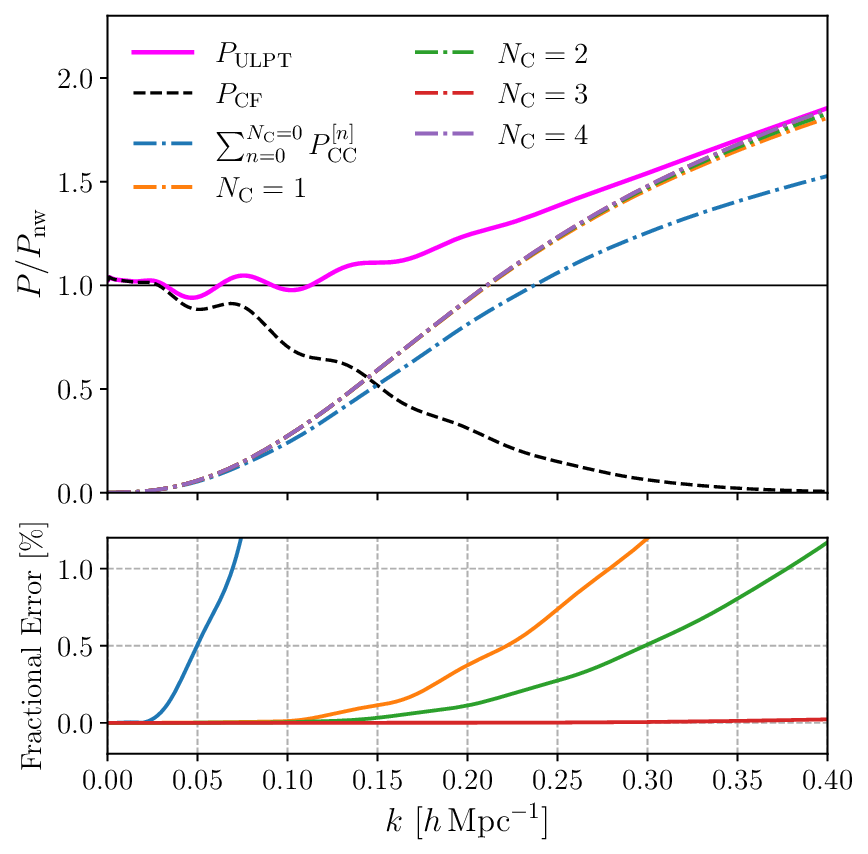}
    \caption{
    One-loop ULPT power spectrum at $z = 0.0$.
    \textit{Upper panel:} The one-loop ULPT power spectrum is decomposed into the convolution-free component, $P_{\rm CF}/P_{\rm nw}$, and the cumulative convolution-containing contributions, $\sum_{n=0}^{N_{\rm C}} P_{\rm CC}^{[n]}/P_{\rm nw}$, for $N_{\rm C} = 0, 1, 2, 3, 4$. Here, $P_{\rm nw}$ denotes the no-wiggle linear power spectrum. The total ULPT power spectrum, which is sufficiently converged and can be regarded as the full result, is defined as $P_{\rm ULPT} = P_{\rm CF} + \sum_{n=0}^{4} P_{\rm CC}^{[n]}$ and is also shown.
    \textit{Lower panel:} Convergence test of the truncated expansion in $P_{\rm CC}$, shown as the fractional deviation from the fully summed result $P_{\rm ULPT}$. The color coding is consistent with the upper panel: $N_{\rm C} = 0$, 1, 2, and 3 correspond to blue, orange, green, and red, respectively. The results demonstrate that the series converges rapidly, achieving sub-percent accuracy for $N_{\rm C} \geq 2$.
    }
    \label{fig:P_CF_CC}
\end{figure}

In this section, we validate the convergence of the expansion method for evaluating the convolution integrals induced by the displacement-mapping factor, as proposed in Sec.~\ref{sec:approx_cov}. Specifically, we perform a detailed analysis at redshift \( z = 0 \), where nonlinear effects are maximized, and identify the number of expansion terms in \( P_{\rm CC} \) required to achieve sufficient accuracy. Since the nonlinear contribution becomes smaller at higher redshifts, convergence confirmed at \( z = 0 \) guarantees even better accuracy at \( z > 0 \), thereby justifying the use of the same approximation scheme at higher redshifts.

As shown in Eq.~\eqref{eq:CF_CC}, the total power spectrum can be decomposed into two distinct components based on whether they involve convolution integrals: the convolution-free term \( P_{\rm CF} \), and the convolution-containing term \( P_{\rm CC} \). The upper panel of Fig.~\ref{fig:P_CF_CC} displays \( P_{\rm CF}/P_{\rm nw} \) and the cumulative sums \( \sum_{n=0}^{N_{\rm C}} P^{[n]}_{\rm CC}/P_{\rm nw} \) for \( N_{\rm C} = 0, 1, 2, 3, 4 \), where \( P_{\rm nw} \) denotes the no-wiggle linear power spectrum without BAO~\cite{Eisenstein:1997ik,Hamann:2010pw,Chudaykin:2020aoj}.

In accordance with Eq.~\eqref{eq:CFCC}, \( P_{\rm CF} \) asymptotically approaches the linear power spectrum on large scales and exhibits exponential suppression on small scales due to the displacement variance. Conversely, \( P_{\rm CC} \) is subdominant on large scales but dominates the nonlinear shape of the power spectrum on small scales. Together, these two components constitute the full one-loop ULPT power spectrum.

Regarding the convergence of the \( P_{\rm CC} \) expansion, we find that truncating at \( N_{\rm C} = 1 \) already captures the majority of the nonlinear correction. As \( N_{\rm C} \) increases to 2, 3, and 4, the contribution from higher-order terms decreases monotonically, indicating good convergence of the series expansion.

To assess this quantitatively, the lower panel of Fig.~\ref{fig:P_CF_CC} shows the relative error with respect to the reference result obtained by summing up to $N_{\rm C}=4$, which can be regarded as sufficiently converged to represent the total ULPT power spectrum:
\begin{equation}
    \text{Error [\%]} = 100 \times \left| \frac{P_{\rm CF} + \sum_{n=0}^{N_{\rm C}} P_{\rm CC}^{[n]} - P_{\rm ULPT}}{P_{\rm ULPT}} \right|,
\end{equation}
where we define \( P_{\rm ULPT} \equiv P_{\rm CF} + \sum_{n=0}^{N_{\rm C}=4} P_{\rm CC}^{[n]} \). The figure demonstrates that for \( N_{\rm C} = 1 \), the deviation reaches about \( 0.4\% \) at \( k = 0.2\,h\,{\rm Mpc}^{-1} \), and exceeds \( 2\% \) at \( k = 0.4\,h\,{\rm Mpc}^{-1} \). Increasing the expansion to \( N_{\rm C} = 2 \), the error reduces to \( 0.1\% \) at \( k = 0.2\,h\,{\rm Mpc}^{-1} \) and to approximately \( 1.2\% \) at \( k = 0.4\,h\,{\rm Mpc}^{-1} \). At \( N_{\rm C} = 3 \), the approximation achieves sub-percent accuracy over the entire range, with a maximum error of only \( 0.025\% \) at \( k = 0.4\,h\,{\rm Mpc}^{-1} \).

Although not shown in the figure, we confirm that the convergence improves further at higher redshifts. For example, at \( z = 0.5 \), the error with \( N_{\rm C} = 2 \) is below \( 0.03\% \) at \( k = 0.2\,h\,{\rm Mpc}^{-1} \) and around \( 0.4\% \) at \( k = 0.4\,h\,{\rm Mpc}^{-1} \). At \( z = 1.0 \), the respective errors decrease further to \( 0.01\% \) and \( 0.13\% \), confirming the robustness of the expansion scheme.

Since current galaxy surveys such as the Dark Energy Spectroscopic Instrument (DESI)~\cite{DESI:2016fyo} typically utilize power spectrum measurements up to \( k \simeq 0.2\,h\,{\rm Mpc}^{-1} \), we conclude that truncating the expansion at \( N_{\rm C} = 2 \) suffices for most practical applications in cosmological data analysis. Nevertheless, to adopt a conservative stance throughout this paper, we present numerical results based on the more stringent choice of \( N_{\rm C} = 3 \).

The full pipeline begins by obtaining the linear power spectrum from the Dark Emulator, followed by evaluating the one-loop correction using FAST-PT as described in Sec.~\ref{sec:FASTONE}, and performing the convolution integrals of the displacement-mapping factor in Sec.~\ref{sec:approx_cov} to compute the final one-loop ULPT power spectrum. When executed on a standard MacBook Pro, the runtime for each stage of the computation is as follows: retrieving the linear power spectrum takes approximately 0.73 seconds, while the FAST-PT evaluation requires about 0.016 seconds. The convolution integrals for the displacement-mapping factor take 1.0 seconds for $N_{\rm C} = 2$, 1.75 seconds for $N_{\rm C} = 3$, and 2.7 seconds for $N_{\rm C} = 4$. Consequently, the total runtime amounts to approximately 1.8 seconds, 2.5 seconds, and 3.4 seconds for $N_{\rm C} = 2$, 3, and 4, respectively. These timings are measured using 450 logarithmically spaced $k$-points in the range $10^{-4}\,h\,\mathrm{Mpc}^{-1} \leq k \leq 100\,h\,\mathrm{Mpc}^{-1}$, which provides sufficient resolution and coverage for computing the two-point correlation function via \texttt{mcfit}. Since a truncation at $N_{\rm C} = 2$ or $3$ is sufficient to achieve sub-percent accuracy in the convolution expansion, we conclude that the full ULPT power spectrum can be computed in approximately 2 seconds with practically negligible loss of precision. This level of computational efficiency highlights the practical applicability of the method for rapid and repeated model evaluations.

\begin{figure*}[!t]
    \centering
    \includegraphics[width=\textwidth]{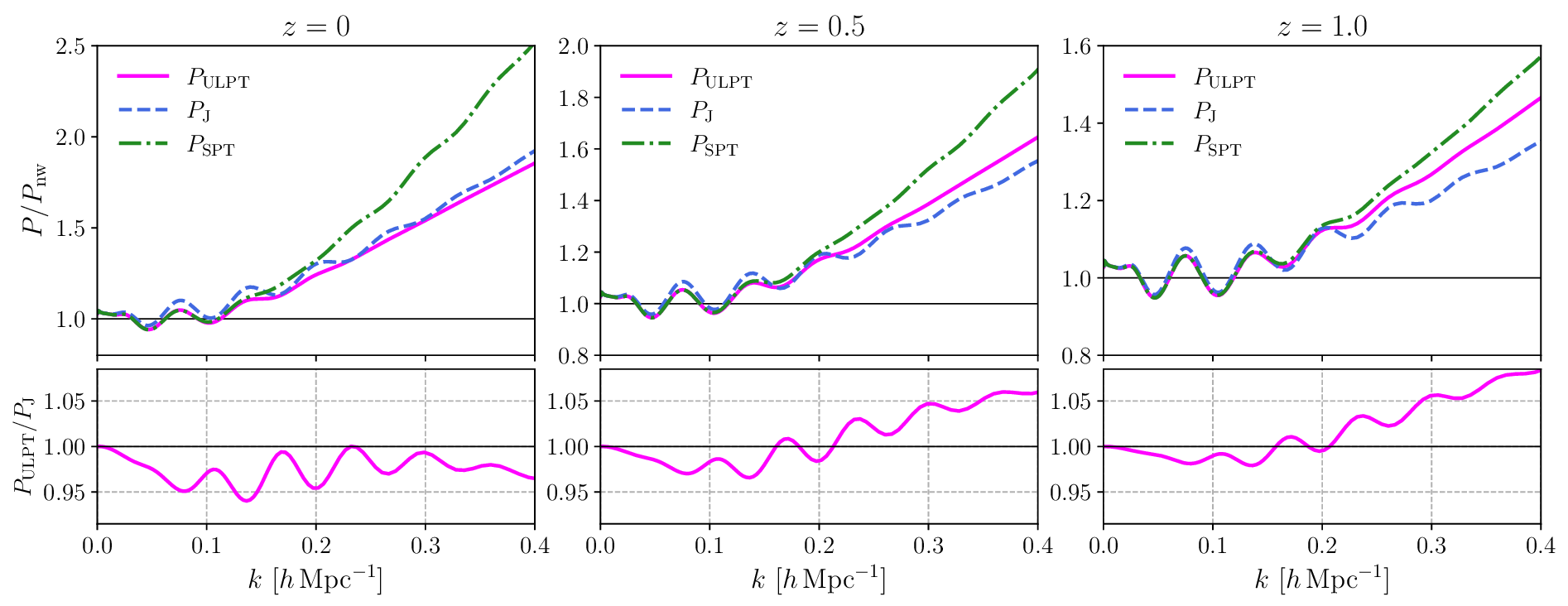}
\caption{
Comparison of the one-loop power spectra at redshifts $z = 0$, $0.5$, and $1.0$.
\textit{Upper panels:} Ratios of the one-loop source power spectrum $P_{\rm J}$, the full ULPT power spectrum $P_{\rm ULPT}$, and the standard SPT power spectrum $P_{\rm SPT}$ to the no-wiggle linear spectrum $P_{\rm nw}$. All spectra include one-loop corrections. The SPT prediction is computed by expanding the displacement-mapping factor to $\mathcal{O}(P_{\rm lin})$, whereas ULPT retains its full exponential form. At $z = 0$ (left panel), $P_{\rm ULPT}$ and $P_{\rm SPT}$ begin to diverge around $k \simeq 0.1\,h\,{\rm Mpc}^{-1}$, with the latter exhibiting excessive nonlinear growth due to the breakdown of the perturbative expansion.
\textit{Lower panels:} Ratios $P_{\rm ULPT}/P_{\rm J}$ at each redshift, illustrating the modulation induced by the displacement-mapping factor. At $z = 0$, the modulation is purely suppressive, reaching 3--5\% up to $k \simeq 0.4\,h\,{\rm Mpc}^{-1}$. At higher redshifts ($z = 0.5$ and $z = 1.0$), the modulation becomes scale-dependent: while a suppression of 2--3\% persists at large scales ($k \lesssim 0.2\,h\,{\rm Mpc}^{-1}$), an enhancement up to 5\% emerges at smaller scales ($k \gtrsim 0.2\,h\,{\rm Mpc}^{-1}$), with the effect being most pronounced at $z = 1.0$. These ratios also exhibit oscillatory features around the BAO scale, reflecting the difference in how $P_{\rm ULPT}$ and $P_{\rm J}$ encode the BAO signal. A more detailed analysis of BAO damping is presented in Fig.~\ref{fig:xi_BAO}.
}
    \label{fig:P_J_ULPT}
\end{figure*}

\section{Source Power Spectrum vs.\ Displacement-Mapping Factor}
\label{sec:source_power}

In this section, we investigate the impact of the displacement-mapping factor on the one-loop power spectrum. Specifically, we compare the one-loop source power spectrum $P_{\rm J}$ with the full one-loop ULPT power spectrum $P_{\rm ULPT}$, which includes convolution integrals arising from the displacement-mapping factor.

For reference, we also compute the one-loop SPT power spectrum $P_{\rm SPT}$. While the computation follows standard procedures implemented in FAST-PT, its structural relationship to the ULPT formulation is clarified in Eq.~\eqref{eq:SPT_J_DM}: both share the same one-loop source power spectrum, but differ in how the displacement-mapping factor is treated. In SPT, this factor is expanded to $\mathcal{O}(P_{\rm lin})$, whereas ULPT retains its full exponential form. Comparing $P_{\rm SPT}$ and $P_{\rm ULPT}$ thus allows us to isolate and quantify the consequences of expanding versus preserving the displacement-mapping factor.

The upper panels of Fig.~\ref{fig:P_J_ULPT} show the ratios $P_{\rm J}/P_{\rm nw}$, $P_{\rm ULPT}/P_{\rm nw}$, and $P_{\rm SPT}/P_{\rm nw}$ at redshifts $z = 0$, $0.5$, and $1.0$, where $P_{\rm nw}$ denotes the no-wiggle linear power spectrum. All spectra include one-loop corrections.

We observe that $P_{\rm ULPT}$ and $P_{\rm SPT}$ begin to diverge around $k = 0.1\,h\,{\rm Mpc}^{-1}$ at redshift $z = 0.0$ (left panel of Fig.~\ref{fig:P_J_ULPT}), where nonlinear effects are most pronounced. As $k$ increases, the SPT prediction grows rapidly and eventually overshoots, while the ULPT result shows a more moderate nonlinear growth. This divergence arises entirely from the treatment of the displacement-mapping factor: both approaches share the same one-loop source term $P_{\rm J}$, but SPT expands the exponential, whereas ULPT retains its full nonperturbative form. The excessive growth of $P_{\rm SPT}$ at small scales reflects the breakdown of the expansion and underscores the importance of preserving the exponential structure.

Comparing $P_{\rm ULPT}$ and $P_{\rm J}$, we find that the overall shape of the ULPT power spectrum is primarily governed by the source term. The displacement-mapping factor modulates its amplitude across a broad range of scales, either suppressing or enhancing it depending on redshift and wavenumber.

To quantify this modulation, the lower panels of Fig.~\ref{fig:P_J_ULPT} plot the ratio $P_{\rm ULPT}/P_{\rm J}$ at redshifts $z = 0$, $0.5$, and $1.0$. The results show that the displacement-mapping factor modifies the source power spectrum by a few percent, with the sign and magnitude of the modulation depending on both scale and redshift. At $z = 0$, the effect is purely suppressive, reaching a level of 3--5\% up to $k \simeq 0.4\,h\,{\rm Mpc}^{-1}$. At higher redshifts ($z = 0.5$ and $z = 1.0$), the modulation exhibits a qualitatively different trend: while a suppression of 2--3\% persists at large scales ($k \lesssim 0.2\,h\,{\rm Mpc}^{-1}$), the displacement-mapping factor enhances the source power spectrum by up to 5\% at smaller scales ($k \gtrsim 0.2\,h\,{\rm Mpc}^{-1}$). This enhancement becomes more pronounced at higher redshift, with the most significant amplification observed at $z = 1.0$. These results indicate a redshift-dependent transition in the displacement-mapping effect, from overall suppression at low redshift to scale-dependent enhancement at high redshift.

Importantly, these ratios exhibit oscillatory features at the BAO scale. This is because $P_{\rm ULPT}$ and $P_{\rm J}$ encode different BAO behavior. In particular, the displacement-mapping factor damps the BAO signal in addition to modulating the broadband shape, resulting in residual oscillations in their ratio.

A more detailed analysis of nonlinear BAO damping will be presented in Sec.~\ref{sec:BAO}.

\section{Simulation-Based Emulators: Dark Emulator \& Euclid Emulator 2}
\label{sec:emu}

\begin{table}[!htbp]
\centering
\caption{Parameter ranges for Dark Emulator and Euclid Emulator 2}
\begin{tabular}{lcc}
\hline
\textbf{Parameter} & \textbf{Dark Emulator} & \textbf{Euclid Emulator 2} \\
\hline
$\omega_b = \Omega_b h^2$ & $[0.02114, 0.02336]$ & $[0.02133, 0.02373]$ \\
$\omega_c = \Omega_c h^2$ & $[0.10782, 0.13178]$ & $[0.11293, 0.15537]$ \\
$\Omega_{\rm de}$ or $\Omega_m$ & $[0.54752, 0.82128]$ (DE) & $[0.23116, 0.38666]$ ($\Omega_m$) \\
$\ln(10^{10} A_s)$ & $[2.4752, 3.7128]$ & $[2.398, 3.178]$ \\
$n_s$ & $[0.9163, 1.0127]$ & $[0.928, 1.000]$ \\
$w_0$ & $[-1.2, -0.8]$ & $[-1.3, -0.5]$ \\
$w_a$ & --- & $[-1.73, 0.5]$ \\
$\sum m_\nu$ [eV] & fixed (0.06 eV) & $[0.0, 0.6]$ \\
$h$ & --- & $[0.645, 0.795]$ \\
\hline
\end{tabular}
\label{tab:param_ranges}
\end{table}

\begin{figure}[!htbp]
    \centering
    \includegraphics[width=\columnwidth]{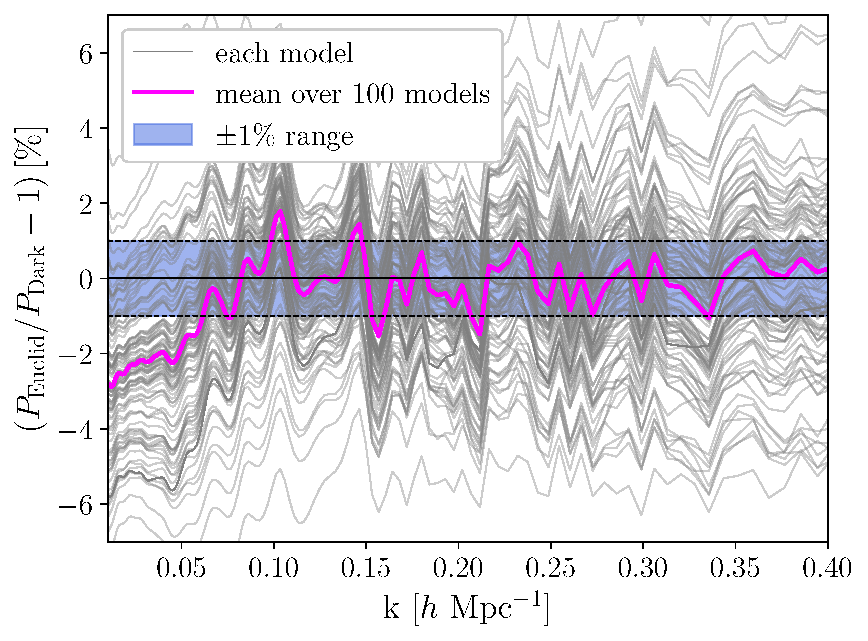}
\caption{
Fractional difference in the nonlinear matter power spectrum between \textit{Euclid Emulator 2} and \textit{Dark Emulator}, evaluated at redshift $z=0$ for 100 randomly sampled cosmological models within the overlapping parameter space of the two emulators. The quantity plotted is $(P_{\mathrm{Euclid}} - P_{\mathrm{Dark}})/P_{\mathrm{Dark}}$. While the deviation exceeds 5\% in some individual models, the mean deviation (magenta line) remains within $\pm 1\%$ for $k \gtrsim 0.05\,h\,\mathrm{Mpc}^{-1}$, confirming the mutual consistency of the two emulators in the nonlinear regime. At larger scales ($k \lesssim 0.05\,h\,\mathrm{Mpc}^{-1}$), the mean deviation can reach up to 2\%, primarily due to sample variance in \textit{Dark Emulator}, as discussed in Sec.~\ref{sec:various}.
}
    \label{fig:Euclid_Dark}
\end{figure}

\subsection{Overview and Motivation}

To evaluate the validity of ULPT predictions across a wide range of cosmological models, we employ simulation-based emulators that provide fast and accurate estimates of the nonlinear matter power spectrum. In this work, we primarily use the \textit{Dark Emulator}~\cite{Nishimichi:2018etk}, which was originally developed to compute halo statistics such as halo--halo and halo--matter power spectra with high efficiency and accuracy. Although Dark Emulator also provides predictions for the nonlinear matter power spectrum, this component has received less validation in the literature because the emulator was designed mainly for halo-related observables.

In the present study, we use only the nonlinear matter power spectrum from the Dark Emulator, without utilizing its halo modules. This choice is made to ensure consistency in future developments, where ULPT may be extended to compute one-loop corrections for halo statistics. Using the same emulator for both matter and halo predictions will allow for seamless validation within a unified framework.

However, given the relatively limited validation of the matter power spectrum component in the Dark Emulator, we perform an independent cross-check using a second emulator, \textit{Euclid Emulator 2}~\cite{Euclid:2020rfv}, which has been carefully validated to achieve sub-percent accuracy across a wide parameter range, including models with massive neutrinos and time-dependent dark energy.

\subsection{Simulation Specifications}
\label{sec:sim_details}

The Dark Emulator is based on the Dark Quest $N$-body simulation suite, which covers 101 flat $w$CDM cosmologies. Each model is simulated with a high-resolution run using $2048^3$ particles in a $2\,h^{-1}\mathrm{Gpc}$ box. To suppress statistical noise, the fiducial Planck cosmology is simulated with 14 independent realizations, enabling a robust estimate of cosmic variance. Accordingly, most of our comparisons between ULPT and emulator predictions are first performed using this fiducial model.

In contrast, Euclid Emulator 2 is trained on 127 cosmological models, each simulated with a paired-and-fixed (P+F) initial condition scheme. For each cosmology, two simulations are performed using $3000^3$ particles in a $1\,h^{-1}\mathrm{Gpc}$ box, resulting in a total of 254 simulations. The P+F method is known to significantly suppress sample variance, particularly on large scales. This design suggests that \textit{Euclid Emulator 2} exhibits inherently lower sample variance than \textit{Dark Emulator} across most relevant cosmological models.

\subsection{Cross-Validation Between Emulators}

To assess the reliability of the Dark Emulator in cosmologies beyond the fiducial model, we perform a direct cross-comparison with Euclid Emulator 2. Using the common parameter region specified in Table~\ref{tab:param_ranges}, we randomly sample 100 cosmological parameter sets that lie within the overlapping validity range of both emulators. For each model, we compute the nonlinear matter power spectrum from both emulators and examine their fractional difference.

Figure~\ref{fig:Euclid_Dark} displays the relative deviation between the two emulator predictions for these 100 cosmological models at redshift $z = 0$, where nonlinear effects are strongest. While some individual models show deviations exceeding 5\% at certain scales, the \textit{mean deviation remains within $\pm 1\%$}, indicating that the two emulators are statistically consistent across the sampled parameter space.

The substantial scatter observed for individual cosmologies is attributed primarily to the larger statistical variance inherent in the Dark Emulator, as anticipated from the simulation specifications in Sec.~\ref{sec:sim_details}. In particular, for scales $k \lesssim 0.05\,h\,\mathrm{Mpc}^{-1}$, even the mean deviation between the two emulators can reach up to 2\%, again reflecting the sample variance in Dark Emulator.

As we demonstrate in Sec.~\ref{sec:various}, the ULPT predictions on large scales agree well with the Euclid Emulator 2 results, confirming that the perturbative framework remains reliable in this regime. 

\begin{figure}[!t]
    \centering
    \includegraphics[width=\columnwidth]{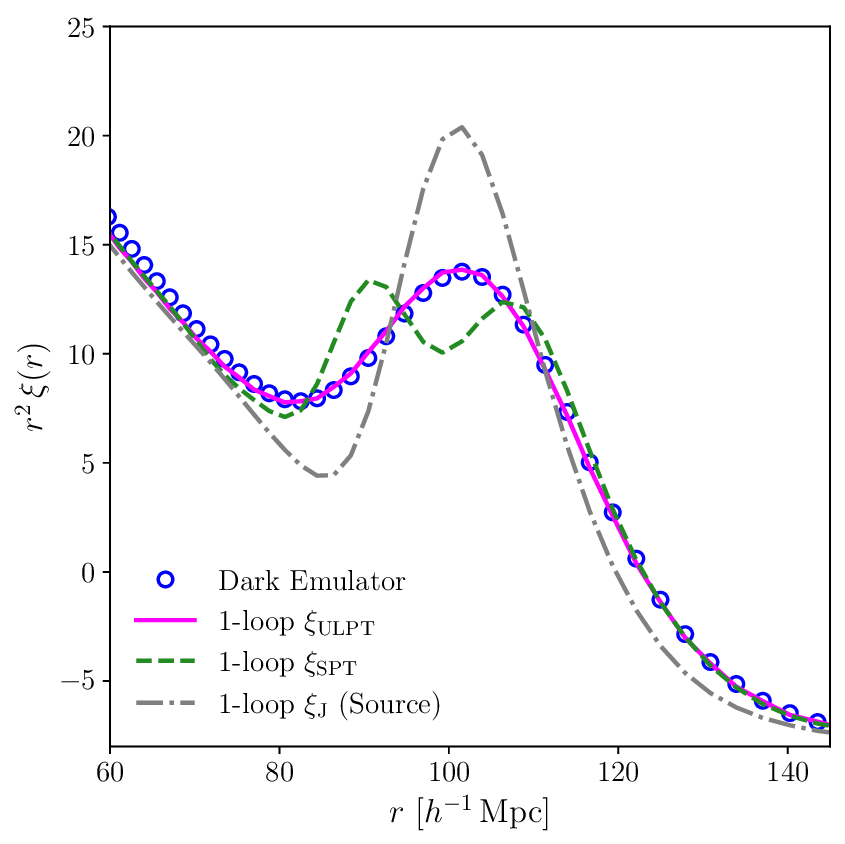}
\caption{
Comparison of the BAO feature in the two-point correlation function at $z = 0$ as predicted by different theoretical models. Shown are the one-loop ULPT correlation function $\xi_{\rm ULPT}$ (magenta solid), the one-loop SPT correlation function $\xi_{\rm SPT}$ (green dashed), and the one-loop source correlation function $\xi_{\rm J}$ (gray dot-dashed), along with the nonlinear correlation function from the \textit{Dark Emulator} (blue circles). The ULPT result incorporates the full exponential form of the displacement-mapping factor, while the SPT result is obtained by expanding this factor to leading order. The source term $\xi_{\rm J}$ does not include any displacement effect. The ULPT prediction closely matches the emulator result and reproduces the nonlinear damping of the BAO peak, whereas the SPT approximation distorts the peak structure.
}
    \label{fig:xi_BAO}
\end{figure}

\section{Nonlinear Damping of BAO in the Correlation Function}
\label{sec:BAO}

In this section, we investigate how ULPT describes the nonlinear damping of the BAO feature in the two-point correlation function. The behavior at smaller scales is deferred to Sec.~\ref{sec:fid}.

\subsection{Displacement Mapping and Nonlinear BAO}

To elucidate the role of the displacement-mapping factor in modeling BAO suppression, we compare several correlation functions at one-loop order. Specifically, Fig.~\ref{fig:xi_BAO} presents four results: (i) the source correlation function $\xi_{\rm J}$, which excludes the displacement-mapping effect; (ii) the one-loop SPT correlation function $\xi_{\rm SPT}$, where the displacement-mapping factor is expanded to leading order; (iii) the ULPT correlation function $\xi_{\rm ULPT}$, which retains the full exponential structure; and (iv) the nonlinear correlation function computed from the \textit{Dark Emulator}. All computations are performed using the fiducial cosmological parameters introduced in Sec.~\ref{sec:intro}, for which the \textit{Dark Emulator} exhibits minimal variance.

This comparison highlights the critical role of the displacement-mapping factor. The source-only result $\xi_{\rm J}$ exhibits no suppression of the BAO feature. The SPT result $\xi_{\rm SPT}$, obtained by expanding the displacement-mapping factor, distorts the shape of the BAO peak. In contrast, the ULPT prediction $\xi_{\rm ULPT}$ accurately reproduces the nonlinear damping effect and shows excellent agreement with the emulator result. These findings emphasize the importance of preserving the full exponential structure of the displacement-mapping factor to correctly capture the nonlinear behavior of BAO.

\subsection{Structure of One-Loop Corrections in ULPT}
\label{sec:structure}

\begin{figure}[!t]
    \centering
    \includegraphics[width=\columnwidth]{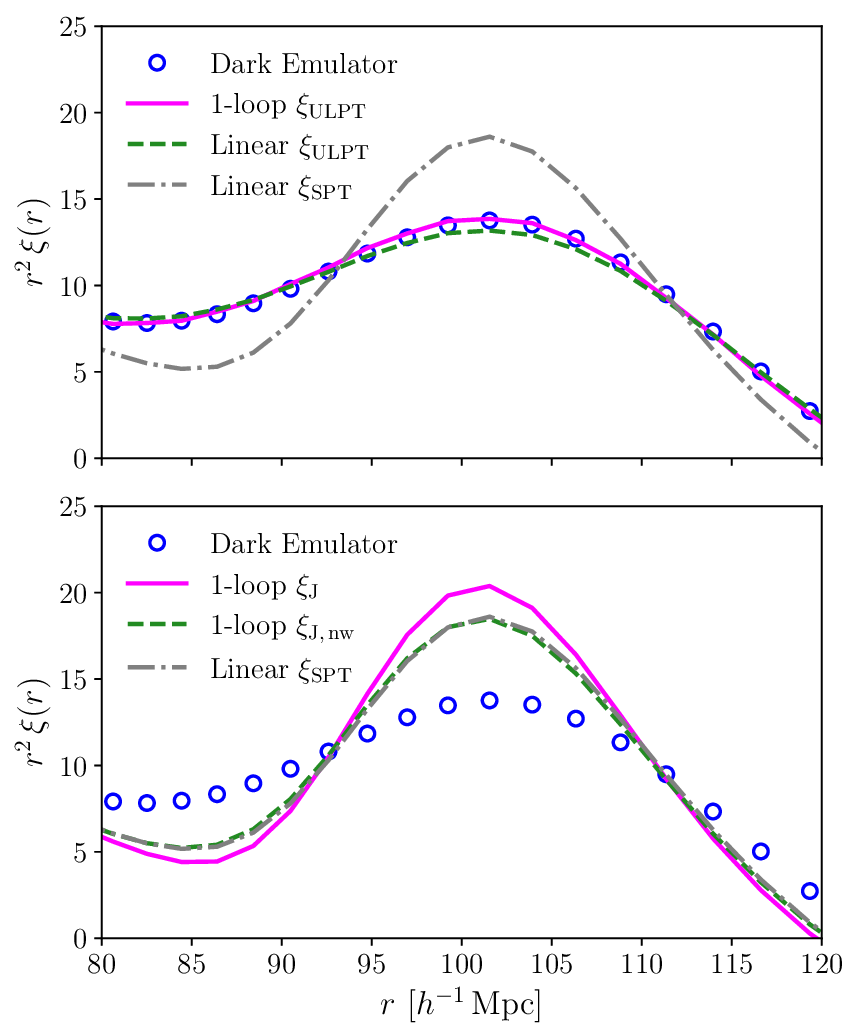}
\caption{
Supplementary comparison of correlation functions to clarify the roles of linear and nonlinear effects in BAO damping. \textbf{Top panel:} Comparison between the linear SPT prediction (gray dot-dashed), the linear ULPT prediction (green dashed), and the full one-loop ULPT result (solid magenta). The linear ULPT result includes the displacement-mapping factor but uses a linear source term. The one-loop correction enhances the BAO peak relative to the linear ULPT result and shows better agreement with the \textit{Dark Emulator} (blue circles). \textbf{Bottom panel:} Comparison between the linear SPT result (gray dot-dashed), the one-loop source correlation function $\xi_{\rm J}$ (magenta solid), and a modified source correlation function $\xi_{\rm J,nw}$ (green dashed), constructed by replacing the linear power spectrum in the 13-type term with the no-wiggle spectrum. The absence of BAO peak enhancement in $\xi_{\rm J,nw}$ indicates that the 13-type term in the source correlation function is responsible for the nonlinear sharpening of the BAO feature.
}
    \label{fig:xi_BAO_suppl}
\end{figure}

To further clarify the internal structure of ULPT, the top panel of Fig.~\ref{fig:xi_BAO_suppl} compares the linear correlation functions from SPT and ULPT with the full one-loop ULPT prediction. The linear ULPT correlation function is computed by evaluating the source correlation function at linear order while fully incorporating the displacement-mapping factor. Although this approximation already captures the general shape of BAO suppression, it slightly underestimates the peak amplitude compared to the one-loop ULPT and emulator results. This discrepancy indicates that one-loop corrections to the source term $\xi_{\rm J}$ mildly enhance the BAO peak.

The bottom panel of Fig.~\ref{fig:xi_BAO_suppl} compares the linear SPT result with the one-loop source correlation function. Since both exclude the displacement-mapping factor, their difference reflects purely the nonlinear evolution contained in $\xi_{\rm J}$. The one-loop source correlation function exhibits a small but noticeable increase in the BAO peak relative to the linear SPT prediction.

As shown in Eq.~\eqref{eq:1loop_J}, the one-loop source power spectrum correction $P_{\rm J}^{(\text{1-loop})}$ consists of a 13-type term and a 22-type term. The 13-type term is proportional to the linear power spectrum and retains the BAO signal, while the 22-type term involves mode-coupling integrals that tend to smear out oscillatory features. Therefore, the enhancement of the BAO peak in $\xi_{\rm J}$ is primarily attributed to the 13-type term.

To gain further insight into the 22-type contribution, we now examine how mode-coupling integrals impact the nonlinear damping of the BAO feature. The BAO signal represents an isotropic oscillation, and its structure can be smoothed through angular averaging inherent in the mode-coupling process. This smoothing arises from destructive interference between oscillatory modes with different phases. However, the efficiency of this phase cancellation depends nontrivially on the detailed form of the linear power spectrum and the specific structure of the coupling kernel. For example, if the SPT kernel $F_2$ is replaced by an arbitrary kernel, it is not guaranteed that the resulting integrals will fully suppress the BAO signal. In practice, for $\Lambda$CDM-like cosmologies, this smoothing is known to occur and provides a subdominant but non-negligible contribution to the total nonlinear power spectrum~\cite{Crocce:2007dt}.

In particular, when the coupling kernel contains scale-dependent terms such as $(k_1/k_2)^2$ or $k_1/k_2$, as listed in Table~\ref{tab:c_alpha_beta_l}, the integrals become sensitive to specific phases in the infrared limit $k_2 \to 0$. This sensitivity weakens the cancellation of BAO oscillations and results in a residual imprint in the final spectrum. Accordingly, such terms in the SPT 22-type contribution $P^{(22)}$ have been used in analytic models to characterize nonlinear BAO damping. The widely used IR-resummed model expresses the power spectrum as~\cite{Baldauf:2015xfa,Blas:2016sfa}
\begin{align}
    P(k) = e^{-k^2\bar{\sigma}_{\rm BAO}^2} P_{\rm w}(k) + P_{\rm nw}(k),
    \label{eq:Plin_IR}
\end{align}
where $P_{\rm w}$ and $P_{\rm nw}$ are the oscillatory (wiggle) and smooth (no-wiggle) components of the linear power spectrum, respectively. The damping scale $\bar{\sigma}_{\rm BAO}^2$ is given by
\begin{align}
    \bar{\sigma}_{\rm BAO}^2 = \frac{1}{3} \int \frac{dk}{2\pi^2} \left[1 - j_0(kr_{\rm BAO}) + 2 j_2(kr_{\rm BAO})\right] P_{\rm lin}(k),
    \label{eq:sigma_BAO}
\end{align}
with $r_{\rm BAO} \sim 105\,h^{-1}\mathrm{Mpc}$ being the characteristic BAO scale. The appearance of spherical Bessel functions $j_0$ and $j_2$ in Eq.~\eqref{eq:sigma_BAO} reflects the $(k_1/k_2)^2$ dependence of the relevant terms in $P^{(22)}$, particularly those corresponding to $(\alpha,\beta,\ell) = (2, -2, 0)$ and $(2, -2, 2)$ in Table~\ref{tab:c_alpha_beta_l}. If all BAO-related contributions in the mode-coupling integrals are neglected, the resulting IR-resummed model reduces to a simplified form that does not involve spherical Bessel functions such as $j_0$ or $j_2$~\cite{Sugiyama:2013gza}.

In the ULPT framework, the 22-type source contribution $P_{\rm J}^{(22)}$ is automatically regularized as shown in Sec.~\ref{sec:source_P}, and terms involving $(k_1/k_2)^2$ do not appear, as shown in Table~\ref{tab:c_alpha_beta_l}. These components are effectively absorbed into the displacement-mapping factor, which encapsulates the dominant nonlinear damping of the BAO. The remaining terms in $P_{\rm J}^{(22)}$ include only those with weaker scale dependence such as $k_1/k_2$ and scale-independent contributions. We thus conclude that the residual BAO signal in $P_{\rm J}^{(22)}$ is clearly subdominant compared to that in the 13-type source contribution $P_{\rm J}^{(13)}$, which is directly proportional to the linear power spectrum and retains the full oscillatory structure. 

To verify this interpretation, we compute a modified version of the source power spectrum in which the linear power spectrum in the 13-type term is replaced by its no-wiggle counterpart:
\begin{equation}
    P_{\rm J,nw}(k)
    = P_{\rm lin}(k) + P_{\rm J}^{(22)}(k) + P_{\rm J}^{(13)}(k)  \left[\frac{P_{\rm nw}(k)}{P_{\rm lin}(k)} \right].
\end{equation}
The corresponding correlation function, denoted the 1-loop $\xi_{\rm J,nw}$, is plotted in the lower panel of Fig.~\ref{fig:xi_BAO_suppl}. This result closely matches the linear SPT prediction and shows no enhancement of the BAO peak, confirming that the 13-type contribution in the source term is essential for capturing the nonlinear sharpening of the BAO feature.

In summary, a complete description of BAO in the context of ULPT requires two complementary effects:
\begin{itemize}
    \item exponential suppression of the BAO amplitude by the displacement-mapping factor;
    \item mild enhancement of the BAO peak due to the 13-type contribution in 
        the 1-loop source power spectrum correction, $P_{\rm J}^{(13)}$.
\end{itemize}

\subsection{IR-Resummed Model Derived from ULPT}
\label{sec:IRresum}

One of the notable features of ULPT is its ability to naturally reproduce the structure of IR-resummed models, which are based on the decomposition of the linear power spectrum into wiggly and no-wiggly components (see, for example, Refs.~\cite{Baldauf:2015xfa,Blas:2016sfa} for standard treatments). In Ref.~\cite{Sugiyama:inprep}, the linear IR-resummed model, Eq.~\eqref{eq:Plin_IR}, was derived directly from the ULPT formulation as the simplest case. In the present work, we generalize this derivation to construct a more comprehensive formulation of IR-resummed models within the ULPT framework. As a concrete example, we derive an IR-resummed expression incorporating one-loop corrections and compare it with the full one-loop prediction obtained directly from ULPT.

Within the ULPT framework, the IR-resummed model is constructed so that the displacement-mapping factor is responsible solely for describing the nonlinear damping of the BAO feature, while the overall shape of the power spectrum is governed by SPT. To proceed, we decompose the source power spectrum into a BAO-carrying wiggle part, $P_{\mathrm{J,w}}$, and a smooth no-wiggle part, $P_{\mathrm{J,nw}}$, as
\begin{equation}
    P_{\mathrm{J}}(k) = P_{\mathrm{J,w}}(k) + P_{\mathrm{J,nw}}(k).
\end{equation}
The corresponding source correlation function is similarly decomposed as
\begin{equation}
    \xi_{\mathrm{J}}(r) = \xi_{\mathrm{J,w}}(r) + \xi_{\mathrm{J,nw}}(r).
\end{equation}
Substituting this decomposition into the general expression for the ULPT power spectrum [Eq.~\eqref{eq:Pk_ULPT}] yields
\begin{align}
    P(k) & = e^{-\overline{\Sigma}(k)} \int d^3 r\, e^{-i\boldsymbol{k} \cdot \boldsymbol{r}}\,
    e^{\Sigma(k,r,\mu)} \xi_{\mathrm{J, w}}(r) \nonumber \\
    & \quad + e^{-\overline{\Sigma}(k)} \int d^3 r\, e^{-i\boldsymbol{k} \cdot \boldsymbol{r}}\,
    e^{\Sigma(k,r,\mu)} \xi_{\mathrm{J, nw}}(r),
\end{align}
where $\mu = \hat{r} \cdot \hat{k}$.

To simplify the convolution integrals arising from the displacement-mapping factor, we introduce an approximation. The wiggle part of the source correlation function $\xi_{\mathrm{J,w}}(r)$ is sharply peaked around the BAO scale and vanishes elsewhere, resembling a delta function. Therefore, we approximate the radial dependence of the displacement-mapping factor by fixing it at $r = r_{\mathrm{BAO}} \sim 105\, h^{-1}\mathrm{Mpc}$, and adopt the conventional choice $\mu = 1$ for the angular dependence. This yields
\begin{align}
    & e^{-\overline{\Sigma}(k)} \int d^3 r\, e^{-i\boldsymbol{k} \cdot \boldsymbol{r}}\,
    e^{\Sigma(k,r,\mu)} \xi_{\mathrm{J, w}}(r) \nonumber \\
    & \quad \approx e^{-\overline{\Sigma}(k)+\Sigma(k,r=r_{\mathrm{BAO}},\mu=1)}\, P_{\mathrm{J,w}}(k),
    \label{eq:P_J_w}
\end{align}
where we used the relation $\int d^3 r\, e^{-i\boldsymbol{k} \cdot \boldsymbol{r}}\, \xi_{\mathrm{J,w}}(r) = P_{\mathrm{J,w}}(k)$.

Next, we consider the no-wiggle component. In the spirit of IR-resummed models, we assume that the displacement-mapping factor primarily affects the BAO feature, while the broadband shape of the power spectrum is governed by SPT. Accordingly, the no-wiggle contribution can be expressed as
\begin{equation}
    P_{\rm SPT,nw}(k) = e^{-\overline{\Sigma}(k)} \int d^3 r\, e^{-i\boldsymbol{k} \cdot \boldsymbol{r}}\,
    e^{\Sigma(k,r,\mu)} \xi_{\mathrm{J,nw}}(r),
    \label{eq:P_J_nw}
\end{equation}
which effectively corresponds to a finite-order SPT prediction evaluated using the no-wiggle source correlation function, and does not contain any BAO oscillations.

Combining the wiggle and no-wiggle results given in Eqs.~\eqref{eq:P_J_w} and \eqref{eq:P_J_nw}, the general form of the IR-resummed model in the ULPT framework is given by
\begin{align}
    P_{\mathrm{IR\text{-}resum}}(k) = e^{-\overline{\Sigma}(k)+\Sigma(k,r=r_{\mathrm{BAO}},\mu=1)}\, P_{\mathrm{J,w}}(k) + P_{\mathrm{SPT,nw}}(k),
    \label{eq:IR_resummed_gene}
\end{align}
where $P_{\mathrm{J,w}}(k)$ captures the BAO oscillatory component modulated by the displacement-mapping factor, and $P_{\mathrm{SPT,nw}}(k)$ corresponds to the broadband shape computed from SPT.

We now derive the explicit form of the IR-resummed model at one-loop order. We begin by evaluating the displacement-mapping exponent in the linear approximation. Using Eq.~\eqref{eq:Sigma_lin_ana}, we obtain
\begin{align}
    & -\overline{\Sigma}^{(\mathrm{lin})}(k) + \Sigma^{(\mathrm{lin})}(k, r = r_{\mathrm{BAO}}, \mu = 1) \nonumber \\
    & \quad = -k^2\left[ \bar{\sigma}^2 - \sigma_0^2(r_{\mathrm{BAO}}) - 2\sigma_2^2(r_{\mathrm{BAO}}) \right] \nonumber \\
    & \quad = -k^2\sigma_{\mathrm{BAO}}^2,
    \label{eq:IR_sigma}
\end{align}
where
\begin{equation}
    \sigma_{\mathrm{BAO}}^2 = \bar{\sigma}^2 - \sigma_0^2(r_{\mathrm{BAO}}) - 2\sigma_2^2(r_{\mathrm{BAO}})
\end{equation}
is the standard BAO damping scale consistent with Eq.~\eqref{eq:sigma_BAO}.

Next, we consider the decomposition of the source power spectrum into wiggle and no-wiggle components. The linear matter power spectrum is split as
\begin{equation}
    P_{\mathrm{lin}}(k) = P_{\mathrm{w}}(k) + P_{\mathrm{nw}}(k).
    \label{eq:wnw}
\end{equation}
The one-loop 13-type source correction $P_{\mathrm{J}}^{(13)}(k)$ is proportional to $P_{\mathrm{lin}}(k)$, and can thus be decomposed using Eq.~\eqref{eq:wnw} as
\begin{equation}
    P_{\mathrm{J}}^{(13)}(k) = \left( \frac{P_{\mathrm{J}}^{(13)}(k)}{P_{\mathrm{lin}}(k)} \right) P_{\mathrm{w}}(k) + \left( \frac{P_{\mathrm{J}}^{(13)}(k)}{P_{\mathrm{lin}}(k)} \right) P_{\mathrm{nw}}(k).
\end{equation}

As discussed in Sec.~\ref{sec:structure}, the BAO contribution from the 22-type source term $P_{\rm J}^{(22)}$ is subdominant and can be neglected in the wiggle part. Therefore, the one-loop wiggle contribution to the source power spectrum is given by
\begin{align}
    P_{\mathrm{J,w}}(k) = P_{\mathrm{w}}(k) + \left[ \frac{P^{(13)}(k) + k^2\bar{\sigma}^2 P_{\mathrm{lin}}(k)}{P_{\mathrm{lin}}(k)} \right] P_{\mathrm{w}}(k),
    \label{eq:PJw}
\end{align}
where we have used the identity $P_{\mathrm{J}}^{(13)} = P^{(13)} + k^2\bar{\sigma}^2 P_{\mathrm{lin}}$ from Eqs.~\eqref{eq:PDM13DM22} and \eqref{eq:P_J13_P_J22}.

On the other hand, the no-wiggle part of the IR-resummed model is obtained by evaluating the SPT prediction up to one-loop order, replacing all occurrences of the linear power spectrum $P_{\mathrm{lin}}$ with the no-wiggle spectrum $P_{\mathrm{nw}}$. The resulting expression reads
\begin{align}
    P_{\mathrm{SPT,nw}}(k) = P_{\mathrm{nw}}(k) + P_{\mathrm{nw}}^{(13)}(k) + P_{\mathrm{nw}}^{(22)}(k),
    \label{eq:PJnw}
\end{align}
where
\begin{align}
    P_{\mathrm{nw}}^{(13)}(k) &= P^{(13)}[P_{\mathrm{lin}} \rightarrow P_{\mathrm{nw}}], \nonumber \\
    P_{\mathrm{nw}}^{(22)}(k) &= P^{(22)}[P_{\mathrm{lin}} \rightarrow P_{\mathrm{nw}}],
\end{align}
Here, $P^{(13)}$ and $P^{(22)}$ represent the one-loop contributions in SPT, and the notation $[P_{\mathrm{lin}} \to P_{\mathrm{nw}}]$ indicates that these terms are evaluated by replacing the linear power spectrum with its no-wiggle counterpart.

Substituting Eqs.~\eqref{eq:IR_sigma}, \eqref{eq:PJw}, and \eqref{eq:PJnw} into the general IR-resummed expression given in Eq.~\eqref{eq:IR_resummed_gene}, we obtain the final one-loop IR-resummed model:
\begin{align}
    P_{\mathrm{IR\text{-}resum}}(k) & = e^{-k^2\sigma_{\mathrm{BAO}}^2} \left[ \left( 1 + k^2\bar{\sigma}^2 \right) + \frac{P^{(13)}(k)}{P_{\mathrm{lin}}(k)} \right] P_{\mathrm{w}}(k) \nonumber \\
    & + P_{\mathrm{nw}}(k) + P_{\mathrm{nw}}^{(13)}(k) + P_{\mathrm{nw}}^{(22)}(k).
\end{align}

\begin{figure}[!t]
    \centering
    \includegraphics[width=\columnwidth]{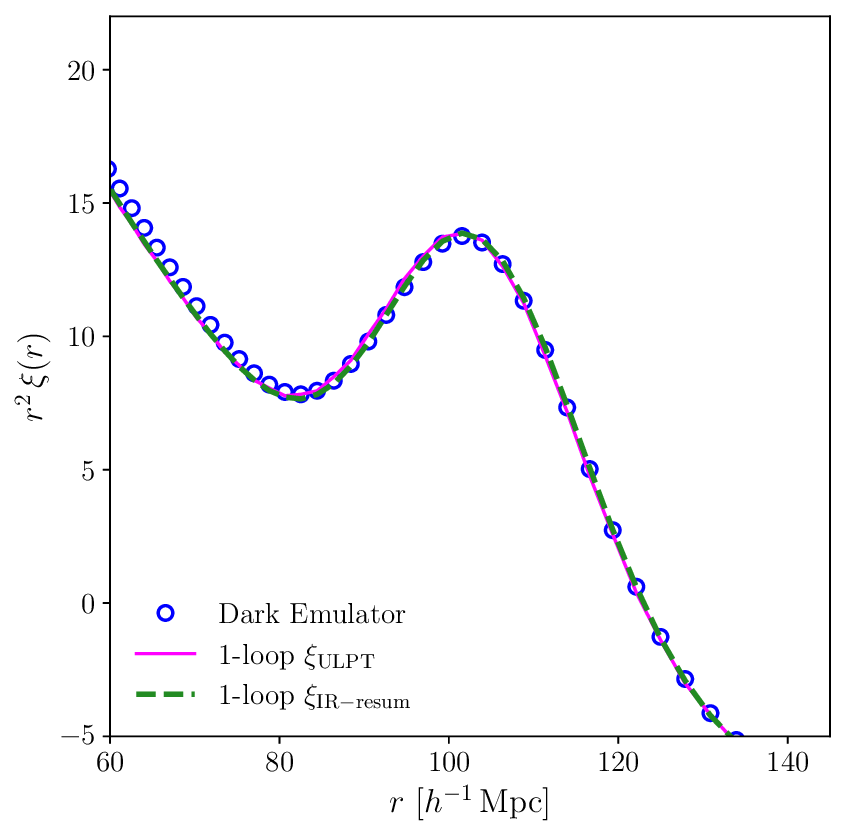}
    \caption{Comparison between the one-loop ULPT prediction and the one-loop IR-resummed model in configuration space. The two models show excellent agreement, demonstrating that the IR-resummed model accurately captures the nonlinear damping of the BAO feature.}
    \label{fig:xi_BAO_IR}
\end{figure}

Figure~\ref{fig:xi_BAO_IR} presents a comparison between the one-loop ULPT prediction and the one-loop IR-resummed model in configuration space. The two results exhibit excellent agreement, confirming that the IR-resummed model accurately reproduces the nonlinear damping of the BAO feature.

\section{Comparison with Emulators}
\label{sec:vs_emu}

We now assess the predictive accuracy and practical applicability of ULPT by comparing its one-loop predictions against simulation-based emulators. In Sec.~\ref{sec:fid}, we examine the fiducial cosmology, for which the Dark Emulator provides high-precision predictions owing to multiple independent realizations. This setting enables a stringent benchmark test of the ULPT framework in both Fourier and configuration space. In Sec.~\ref{sec:various}, we extend the comparison across a broader cosmological parameter space, using 100 randomly sampled models within the overlapping validity range of the Dark Emulator and Euclid Emulator 2. This analysis allows us to evaluate the robustness of ULPT under realistic parameter variations relevant for cosmological inference.

\begin{figure}[!htbp]
    \centering
    \includegraphics[width=\columnwidth]{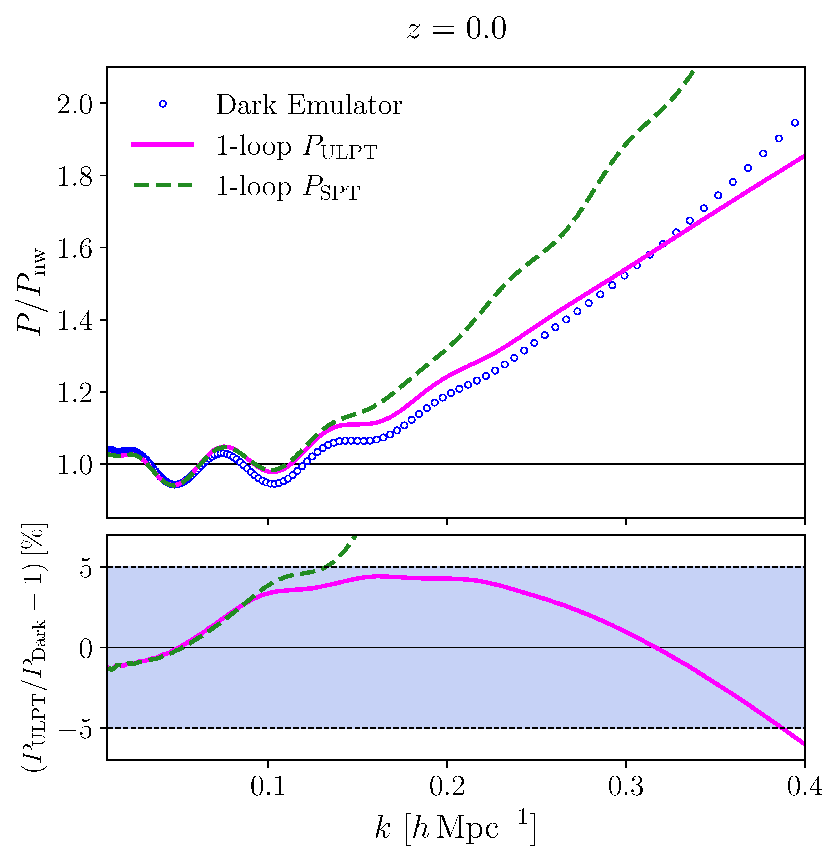}
\caption{
Comparison of the nonlinear matter power spectrum at redshift $z = 0$ for the fiducial cosmology. The upper panel shows the ratio $P(k)/P_{\rm nw}(k)$, where $P_{\rm nw}$ is the no-wiggle linear power spectrum. Predictions from the \textit{Dark Emulator} (blue circles), one-loop ULPT (magenta), and one-loop SPT (green) are shown. The lower panel displays the fractional differences of ULPT and SPT relative to the emulator. All models agree well on large scales ($k \lesssim 0.1\,h\,\mathrm{Mpc}^{-1}$), but SPT deviates significantly on smaller scales, while ULPT maintains agreement within 5\% up to $k \simeq 0.4\,h\,\mathrm{Mpc}^{-1}$.
}
    \label{fig:power_z0}
    \includegraphics[width=\columnwidth]{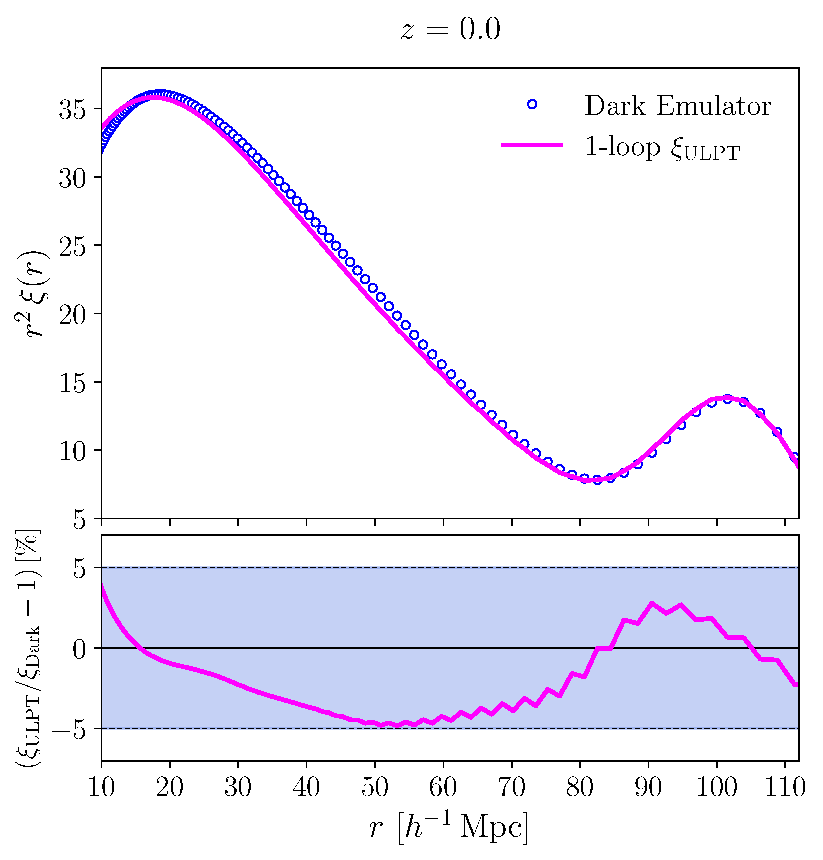}
\caption{
Two-point correlation functions at $z = 0$, obtained by inverse Hankel transforming the power spectra shown in Fig.~\ref{fig:power_z0}. The blue circles represent the \textit{Dark Emulator} prediction, and the magenta line shows the ULPT result. The bottom panel displays the fractional difference between ULPT and the emulator. ULPT accurately reproduces the BAO peak at $r \approx 105\,h^{-1}\mathrm{Mpc}$ and agrees with the emulator to within 5\% down to $r \simeq 10\,h^{-1}\mathrm{Mpc}$.
}

    \label{fig:xi_z0}
\end{figure}

\begin{figure*}[!htbp]
    \centering
    \includegraphics[width=\textwidth]{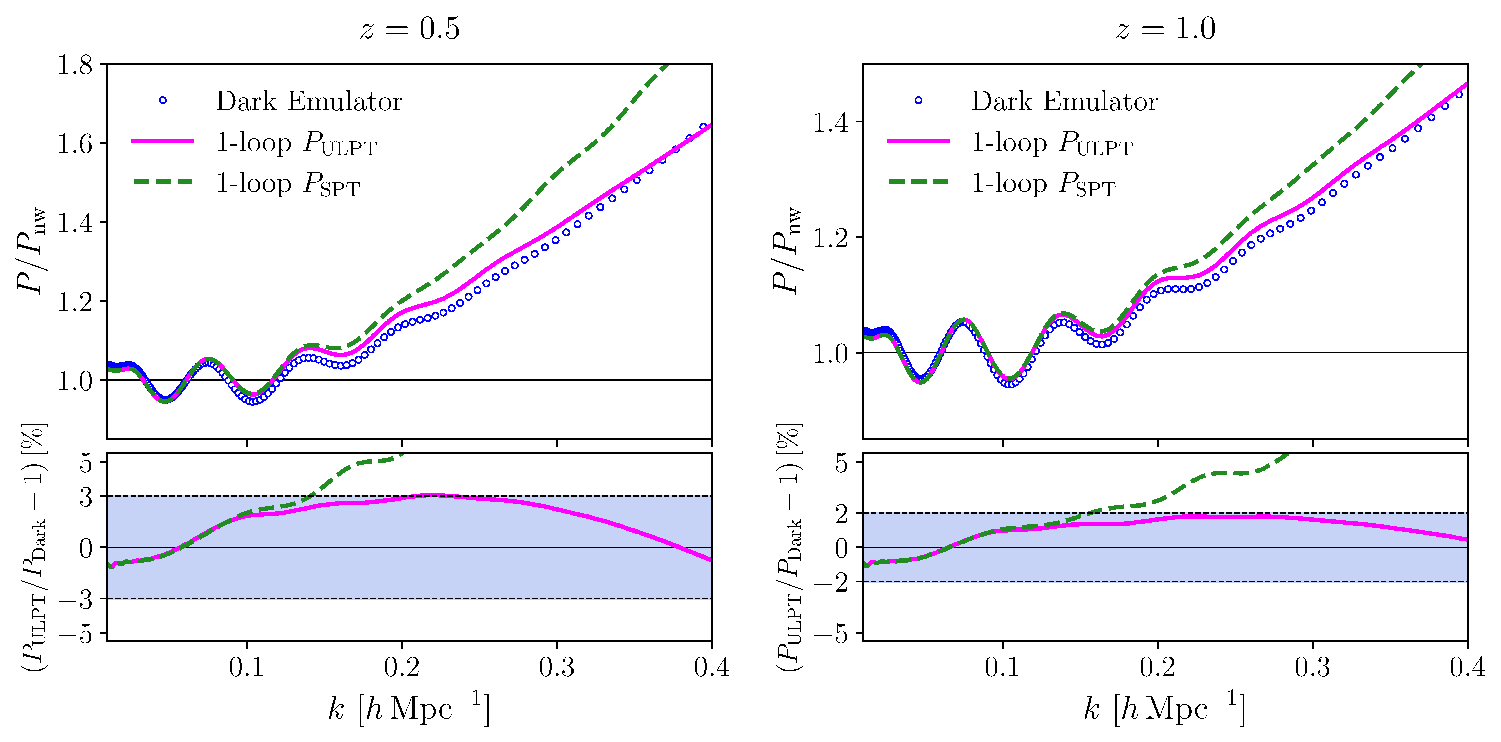}
\caption{
Same as Fig.~\ref{fig:power_z0}, but for redshifts $z = 0.5$ (left) and $z = 1.0$ (right).
ULPT agrees with the \textit{Dark Emulator} to within 3\% at $z = 0.5$ and 2\% at $z = 1.0$, up to $k \simeq 0.4\,h\,\mathrm{Mpc}^{-1}$.
}
    \label{fig:power_z05_z1}
    \vspace{1cm}
    \includegraphics[width=\textwidth]{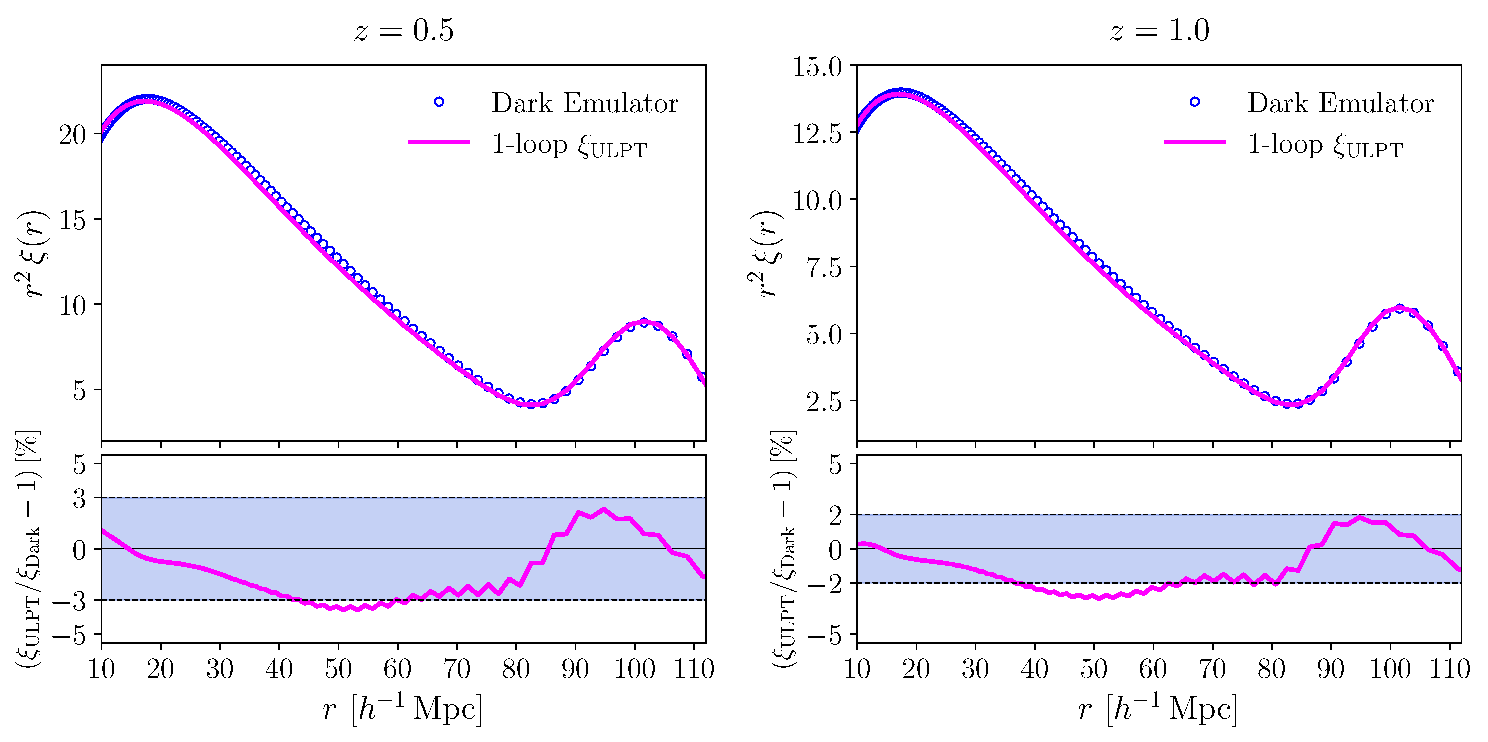}
\caption{
Same as Fig.~\ref{fig:xi_z0}, but for redshifts $z = 0.5$ (left) and $z = 1.0$ (right).
ULPT agrees with the \textit{Dark Emulator} to within 3\% at $z = 0.5$ and 2\% at $z = 1.0$ for $r \gtrsim 10\,h^{-1}\mathrm{Mpc}$, with a mild ($\lesssim 0.5\%$) excess near $r \approx 50\,h^{-1}\mathrm{Mpc}$.
}
    \label{fig:xi_z05_z1}
\end{figure*}

\subsection{Fiducial Cosmology}
\label{sec:fid}

We begin by examining the fiducial cosmology adopted in the Dark Emulator suite, corresponding to the Planck 2015 best-fit $\Lambda$CDM model. This model has been simulated using 14 independent $N$-body realizations, substantially reducing sample variance and yielding reliable emulator predictions for benchmarking theoretical models.

In this subsection, we compare the ULPT prediction for the nonlinear matter power spectrum with the Dark Emulator output, and also present the corresponding two-point correlation function obtained via inverse Hankel transformation.

\subsubsection{$z=0$}

Figure~\ref{fig:power_z0} presents a three-way comparison between the nonlinear matter power spectrum predicted by the \textit{Dark Emulator} ($P_{\rm Dark}$), the one-loop SPT prediction ($P_{\rm SPT}$), and the one-loop ULPT prediction ($P_{\rm ULPT}$) at redshift $z=0$. In the upper panel, each spectrum is divided by the corresponding no-wiggle power spectrum ($P_{\rm nw}$), while the lower panel shows the fractional differences of the perturbative models relative to $P_{\rm Dark}$.

On large scales ($k \lesssim 0.1\,h\,{\rm Mpc}^{-1}$), SPT and ULPT both reproduce the emulator within approximately three percent, indicating that linear growth and next leading-order corrections are accurately captured where bulk displacements remain small.

At small scales ($0.1 \lesssim k \lesssim 0.4\,h\,{\rm Mpc}^{-1}$), the SPT prediction deviates rapidly from the emulator, while ULPT remains accurate to within five percent. This behavior supports the analysis in Sec.~\ref{sec:source_power}, where it was shown that the primary source of inaccuracy in SPT arises from the expansion of the displacement-mapping factor, which is avoided in ULPT by retaining its exponential form.

Configuration-space results are shown in Fig.~\ref{fig:xi_z0}, where the two-point correlation functions obtained via inverse Hankel transforms of $P_{\rm ULPT}$ and $P_{\rm Dark}$ are compared. ULPT agrees with the emulator to within five percent down to $r \simeq 10\,h^{-1}{\rm Mpc}$ and reproduces both the position and amplitude of the BAO peak at $r \approx 105\,h^{-1}{\rm Mpc}$ with sub-percent precision. This level of agreement in real space mirrors the consistency observed in Fourier space, thereby confirming the internal coherence of the ULPT framework.

\subsubsection{$z=0.5$ and $z=1.0$}

Because ongoing galaxy surveys such as DESI and the Prime Focus Spectrograph (PFS)~\cite{PFSTeam:2012fqu} target the redshift range $z \gtrsim 0.5$, Figures~\ref{fig:power_z05_z1} and \ref{fig:xi_z05_z1} extend the comparison to $z=0.5$ and $z=1.0$. For the power spectrum (Fig.~\ref{fig:power_z05_z1}), ULPT agrees with the emulator to within three percent at $z=0.5$ and within two percent at $z=1.0$ up to $k \simeq 0.4\,h\,{\rm Mpc}^{-1}$. This progressive improvement with redshift is a natural consequence of the reduced nonlinear gravitational effects at earlier cosmic times.

The correlation functions shown in Fig.~\ref{fig:xi_z05_z1} exhibit a similarly close agreement: ULPT matches the emulator to better than three percent at $z=0.5$ and two percent at $z=1.0$ for separations $r \gtrsim 10\,h^{-1}\mathrm{Mpc}$. A mild ($\lesssim 0.5\%$) excess appears near $r \approx 50\,h^{-1}\mathrm{Mpc}$, but the overall agreement remains consistent with that seen in the power spectrum comparison.

\subsection{Various Cosmologies}
\label{sec:various}

We next investigate the performance of ULPT across a diverse set of cosmological models. Specifically, we compare the ULPT predictions for the nonlinear matter power spectrum against those of the Dark Emulator and Euclid Emulator 2, using a set of $100$ cosmologies randomly sampled from the parameter region where both emulators are valid. This test is particularly relevant for applications such as Markov Chain Monte Carlo (MCMC) analyses, where repeated evaluations of the power spectrum across a wide parameter space are required.

\begin{figure*}[!htbp]
    \centering
    \includegraphics[width=\textwidth]{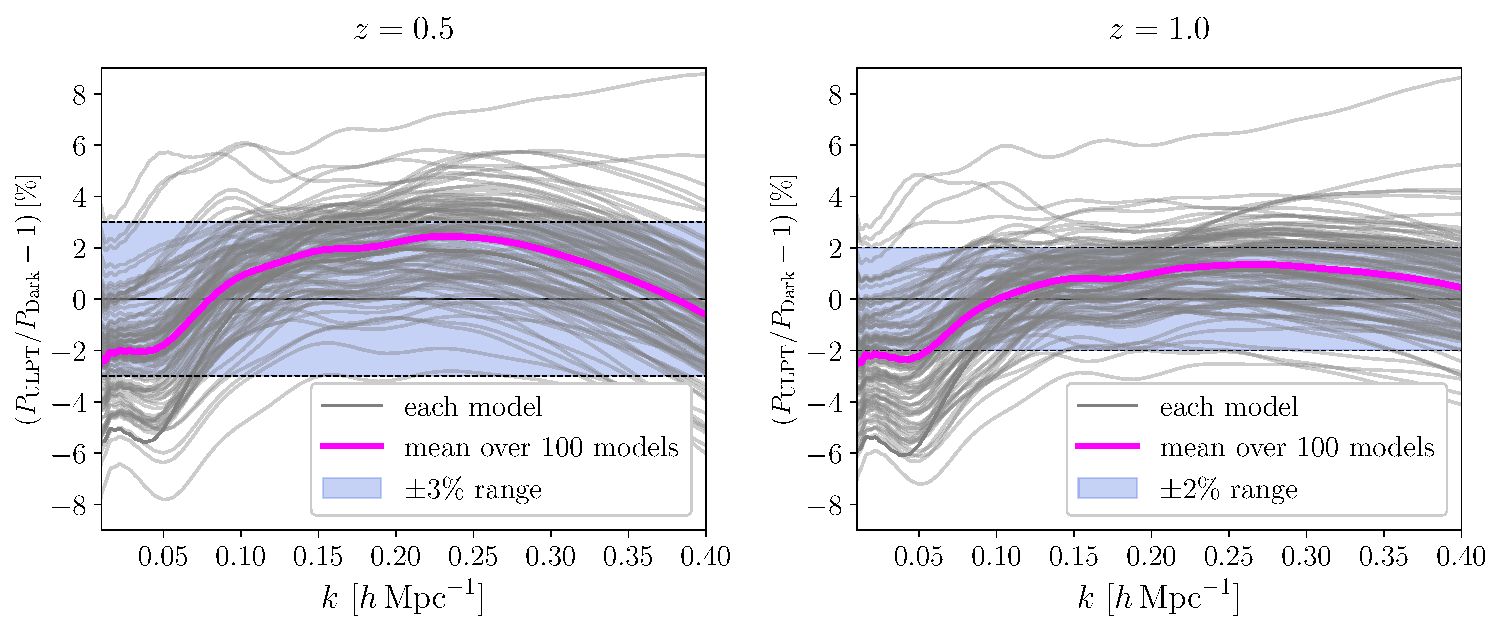}
\caption{
Fractional difference between ULPT and the Dark Emulator for 100 cosmological models at redshifts $z = 0.5$ (left) and $z = 1.0$ (right). Each gray line corresponds to an individual model, and the magenta curve shows the ensemble average. The agreement remains within 3\% at $z = 0.5$ and 2\% at $z = 1.0$, up to $k \simeq 0.4\,h\,\mathrm{Mpc}^{-1}$. The visible scatter, along with the $\sim2\%$ systematic excess of the Dark Emulator prediction relative to ULPT at large scales ($k \lesssim 0.05\,h\,\mathrm{Mpc}^{-1}$), reflects sample variance inherent to the emulator.
}
    \label{fig:ULPT_vs_Emu}
    \vspace{3cm}
    \centering
    \includegraphics[width=\textwidth]{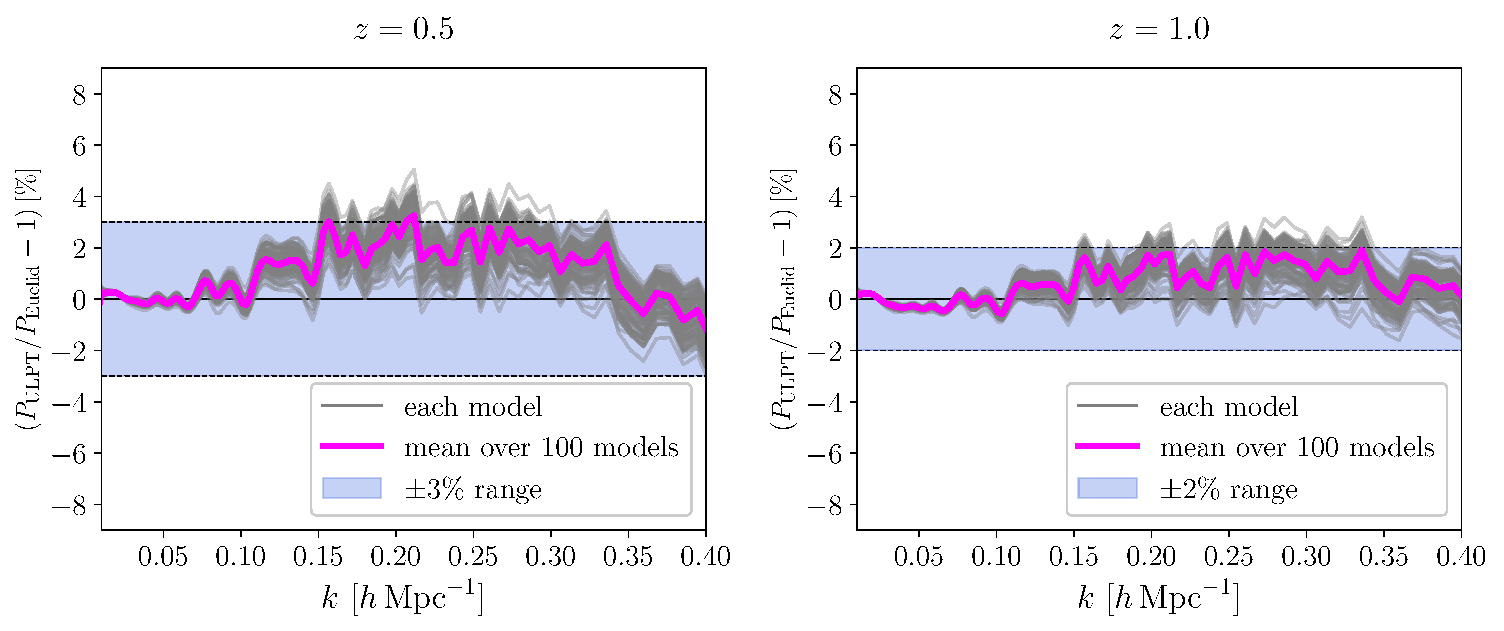}
\caption{
Same as Fig.~\ref{fig:ULPT_vs_Emu}, but comparing ULPT to \textit{Euclid Emulator 2} instead of the Dark Emulator.
The agreement improves at large scales ($k \lesssim 0.05\,h\,\mathrm{Mpc}^{-1}$), where no systematic offset is observed.
The mean difference (magenta) remains within $3\%$ at $z = 0.5$ and $2\%$ at $z = 1.0$, up to $k \simeq 0.4\,h\,\mathrm{Mpc}^{-1}$, demonstrating consistency between ULPT and \textit{Euclid Emulator 2} across the sampled parameter space.
}

    \label{fig:ULPT_vs_Emu_Euclid}
\end{figure*}

Figures~\ref{fig:ULPT_vs_Emu} and \ref{fig:ULPT_vs_Emu_Euclid} present the fractional deviation, $(P_{\rm ULPT}/P_{\rm emu}) - 1$, for each of the sampled cosmologies at redshifts $z = 0.5$ and $z = 1.0$.

\paragraph*{Dark Emulator (Fig.~\ref{fig:ULPT_vs_Emu}).}

At both redshifts, the magenta curves show the mean difference, while the grey lines represent individual models. The mean agreement remains within $\sim3\%$ at $z=0.5$ and $2\%$ at $z=1.0$ up to $k\simeq0.4\,h\,\mathrm{Mpc}^{-1}$, matching the precision already achieved for the fiducial model in Fig.~\ref{fig:power_z05_z1}. However, the individual Dark Emulator realizations display considerable scatter, with deviations exceeding $5\%$ in some cases, due to sample variance (see also Fig.~\ref{fig:Euclid_Dark}). A systematic $\sim2\%$ excess of the Dark Emulator prediction relative to ULPT is also visible at very large scales ($k\lesssim0.05\,h\,\mathrm{Mpc}^{-1}$), and is attributed to the same sample variance effect.

Given that ULPT provides a variance-free theoretical prediction with 2--3\% accuracy across the full range of scales, it serves as a more stable and reliable baseline than the individual emulator outputs, especially in parameter or scale regimes where sample variance is non-negligible.

\paragraph*{Euclid Emulator 2 (Fig.~\ref{fig:ULPT_vs_Emu_Euclid}).}

Replacing the Dark Emulator with Euclid Emulator 2 removes the large-scale bias: no systematic offset is observed for $k\lesssim0.05\,h\,\mathrm{Mpc}^{-1}$, and the mean curve closely follows the ULPT prediction. This excellent agreement on large scales is expected, as perturbation theory is most reliable in this regime. It confirms the accuracy of Euclid Emulator 2 and, reciprocally, supports the validity of the ULPT framework. The ULPT predictions shown here are computed using the same pipeline as in Fig.~\ref{fig:ULPT_vs_Emu}, with the linear matter power spectrum consistently taken from the Dark Emulator.

Owing to its paired-and-fixed simulation strategy, Euclid Emulator 2 exhibits a typical variance of $\lesssim1\%$ across models. After averaging, its predictions remain consistent with those of the Dark Emulator, and the agreement with ULPT again stays within $3\%$ at $z=0.5$ and $2\%$ at $z=1.0$ up to $k\simeq0.4\,h\,\mathrm{Mpc}^{-1}$.

\paragraph*{Implications for parameter inference.}

Because modern MCMC analyses typically require $\mathcal{O}(10^{5})$ evaluations of the matter power spectrum, a fast, variance-free, and percent-level accurate theory is highly desirable. The consistency demonstrated across $100$ cosmological models shows that ULPT meets these requirements while remaining computationally competitive with emulator-based approaches. 

These findings, summarized in Figs.~\ref{fig:ULPT_vs_Emu} and \ref{fig:ULPT_vs_Emu_Euclid}, demonstrate that ULPT enables accurate and efficient cosmological parameter estimation across a broad parameter space.

\section{Future Prospects}
\label{sec:future}

The results presented in this work establish one-loop ULPT as a fast and percent-level accurate description of the nonlinear matter power spectrum across the six-dimensional $w$CDM parameter space supported by current $N$-body emulators. Several extensions naturally follow.

First, since ULPT contains no phenomenological fitting parameters, it can be applied beyond the emulator training volume, offering reliable predictions in cosmologies not covered by simulations. The formalism also admits straightforward extensions to redshift-space distortions and density-field reconstruction, as both effects preserve the structural decomposition into Jacobian and displacement-mapping components.

Incorporating galaxy bias is another promising direction. The original ULPT framework~\cite{Sugiyama:inprep} allows for a bias expansion that avoids renormalization by including additive Galileon operators in the Jacobian deviation. Extending the current implementation to biased tracers will enable predictions for halo clustering statistics that can be directly compared to emulator-based models.

Finally, extending ULPT to two loops is expected to further improve accuracy toward the sub-percent level. Since the residual $2$--$3\%$ discrepancies observed at one loop match the typical size of two-loop corrections, such an extension will provide a crucial test of the framework’s robustness and its ability to surpass the limitations of standard perturbation theory.

\section{Conclusion}
\label{sec:conclusion}

In this work, we have presented a fast and accurate formulation for computing the nonlinear matter power spectrum at one-loop order within the framework of Unified Lagrangian Perturbation Theory (ULPT). The key feature of ULPT is the structural decomposition of the density field into two physically distinct components: the Jacobian deviation, which describes intrinsic nonlinear growth, and the displacement-mapping effect, which captures large-scale convective distortions. 

In this framework, the power spectrum naturally separates into two contributions: a source term that encodes correlations between the displacement field and the Jacobian deviation, and a displacement-mapping factor that encapsulates the uncorrelated impact of large-scale flows. This structural separation not only facilitates physical interpretation but also ensures infrared (IR) safety by construction, playing a central role in the resummation of long-wavelength modes.

We have demonstrated that the breakdown of standard perturbation theory (SPT) at small scales can be traced to its expansion of the displacement-mapping factor, which inherently possesses an exponential form. In contrast, ULPT retains this structure without expansion, leading to improved convergence and a more physically faithful description of nonlinear evolution. Notably, ULPT and SPT share the same one-loop source term \( P_{\rm J}^{(\text{1-loop})} \), but differ in their treatment of the displacement-mapping contribution: SPT expands it, while ULPT resums it. This distinction accounts for the superior performance of ULPT at nonlinear scales.

We have implemented ULPT numerically using a combination of FFTLog and FAST-PT techniques, achieving rapid evaluations suitable for cosmological applications; the full one-loop power spectrum can be computed in approximately 2 seconds on a standard laptop. The predictive accuracy of ULPT has been validated against both the Dark Emulator and Euclid Emulator 2 across 100 cosmological models. In particular, for redshifts \( z \geq 0.5 \), ULPT predictions match simulation-based emulators at the 2--3\% level up to \( k \simeq 0.4\,h\,\mathrm{Mpc}^{-1} \), without invoking any nuisance parameters. Comparable accuracy is also observed in configuration space, where the two-point correlation function remains consistent with emulator results down to \( r \simeq 10\,h^{-1}\mathrm{Mpc} \).

Our analysis also elucidates the mechanism responsible for nonlinear baryon acoustic oscillation (BAO) damping. In ULPT, the displacement-mapping factor induces an exponential suppression of the BAO amplitude by smearing out oscillatory modes, while the nonlinear source term, particularly the 13-type component, mildly sharpens the peak through constructive growth. This complementary interplay reproduces the characteristic BAO features observed in simulations. In contrast, SPT distorts the BAO signal because its expansion of the displacement-mapping factor fails to capture the correct nonlinear structure. This again highlights the importance of preserving the full exponential form, not only for the broadband shape of the power spectrum, but also for accurately describing the nonlinear evolution of the BAO peak.

Looking ahead, ULPT provides a solid theoretical foundation for several future developments. These include consistent extensions to redshift-space distortions, density-field reconstruction, and galaxy bias, all of which can be seamlessly incorporated into the existing formalism. Since ULPT contains no phenomenological fitting parameters, it can also be applied beyond the emulator training volume, offering reliable predictions in cosmologies not covered by simulations. Furthermore, extending ULPT to two-loop order is a natural next step toward achieving sub-percent theoretical precision and may offer new insights into the limitations of standard perturbative approaches.

\begin{acknowledgments}
N.S. acknowledges financial support from JSPS KAKENHI Grant No. 25K07343, administratively hosted by the National Astronomical Observatory of Japan. The author is grateful to the developers of the publicly available Python packages used in this work, including \texttt{mcfit}~\cite{mcfit}, which was used for Hankel transforms, and \texttt{FAST-PT}~\cite{McEwen:2016fjn,Fang:2016wcf}, which enabled efficient evaluation of one-loop corrections. Nonlinear reference spectra were obtained from the \textit{Dark Emulator}~\cite{Nishimichi:2018etk} and the \textit{Euclid Emulator 2}~\cite{Euclid:2020rfv}. N.S. also acknowledges the use of \textit{ChatGPT} (OpenAI) for assistance in language refinement, literature exploration, and Python scripting during the preparation of this manuscript. The author is thankful to the referee for their constructive comments, which have significantly improved the quality and clarity of the paper.
\end{acknowledgments}

% The \nocite command causes all entries in a bibliography to be printed out
% whether or not they are actually referenced in the text. This is appropriate
% for the sample file to show the different styles of references, but authors
% most likely will not want to use it.
%\nocite{*}
\bibliography{ms}% Produces the bibliography via BibTeX.

\end{document}